\documentclass[twocolumn]{aastex6}
\usepackage{graphicx,epstopdf}
\usepackage{subfigure}
\usepackage{rotating}

\usepackage{txfonts}
\usepackage{natbib}
\shorttitle{Atmospheres of Exoplanets}
\shortauthors{Deming, Louie, \& Sheets}

\begin{document}

\title{How to Characterize the Atmosphere of a Transiting Exoplanet}

%
\author{Drake Deming,\altaffilmark{1,2} Dana Louie,\altaffilmark{1,5} \& Holly Sheets\altaffilmark{3,4} }

\altaffiltext{1}{Department of Astronomy, University of Maryland at College Park, College Park, MD 20742, USA}
\altaffiltext{2}{NASA Astrobiology Institute's Virtual Planetary Laboratory}
\altaffiltext{3}{Institute for Research on Exoplanets, Department of Physics, McGill University, PO Box 6128 Centre-Ville STN, Montreal, QC, H3C 3J7, CAN}
\altaffiltext{4}{Present address: Department of Physics, Albion College, 611 E. Porter St., Albion, MI 49224, USA}
\altaffiltext{5}{NASA Earth and Space Sciences Graduate Fellow}

\begin{abstract}
This tutorial is an introduction to techniques used to characterize the atmospheres of transiting exoplanets. We intend it to be a useful guide for the undergraduate, graduate student, or postdoctoral scholar who wants to begin research in this field, but who has no prior experience with transiting exoplanets.  We begin with a discussion of the properties of exoplanetary systems that allow us to measure exoplanetary spectra, and the principles that underlie transit techniques. Subsequently, we discuss the most favorable wavelengths for observing, and explain the specific techniques of secondary eclipses and eclipse mapping, phase curves, transit spectroscopy, and convolution with spectral templates. Our discussion includes factors that affect the data acquisition, and also a separate discussion of how the results are interpreted.  Other important topics that we cover include statistical methods to characterize atmospheres such as stacking, and the effects of stellar activity.  We conclude by projecting the future utility of large-aperture observatories such as the James Webb Space Telescope and the forthcoming generation of extremely large ground-based telescopes. 
\end{abstract}

\keywords{  }



\section{Introduction} \label{sec:intro} 

Transits of extrasolar planets allow us to probe their atmospheres.  While transit techniques are powerful, the measurements and interpretation are by no means easy or routine.  The purpose of this tutorial is to explain the principles and best practices that have been developed by the community to make these measurements.  We also discuss the advantages and limitations of the techniques, and prospects for the future.  We intend this to be a useful guide for the undergraduate, graduate student, or postdoctoral scholar who wants to begin research in this field, but who has no prior experience with transiting planets.  We also hope that this tutorial will be valuable to experienced transit researchers, perhaps serving to remind them of some key principles and results to date. 

This tutorial is organized as follows. Section \ref{sec:properties} articulates the properties of exoplanetary systems that allow us to measure the exoplanetary spectra, and discusses the most favorable wavelengths for observing (Section~\ref{subsec:wavelength}).  Techniques used in the measurements (secondary eclipses, phase curves, transit spectroscopy, eclipse mapping, and convolution methods) are explained in Sections~\ref{subsec:eclipses} through~\ref{subsec:convol}. Section \ref{sec:photspec} compares exoplanetary measurements using the contrasting techniques of photometry and spectroscopy, and Section \ref{sec:groundspace} discusses the advantages of ground-based versus space-borne measurements.  Section \ref{sec:interpreting} explains how the measurements are currently interpreted, including measurements of individual planets, as well as statistical methods.  Activity on the host star can complicate the characterization of a planet's atmosphere, and that is discussed in Section~\ref{sec:activity}.  We conclude in Section~\ref{sec:future} by projecting the capability of future large aperture telescopes in space and on the ground.

Although we mention some important results in this paper, we do not intend this to be a review in the usual sense.  We focus on techniques and methods, rather than results {\it per se}.  Reviews of results in transiting exoplanetary science are given by \citet{seager10}, \citet{marley13}, \citet{burrows14}, \citet{cowan15}, \citet{heng15}, \citet{crossfield15}, \citet{madhu16}, \citet{deming_seager17}, and \citet{fortney18}.  Moreover, the scope of this tutorial is focused on optical and infrared techniques, and not on other important topics such as observations in the ultraviolet or at radio wavelengths. Also, we here focus on transiting planets, but we point out that results on high-contrast imaging of exoplanets are reviewed by \citet{bowler16}.

\section{Properties of Transits} \label{sec:properties}

\begin{figure}
\includegraphics[width=3in]{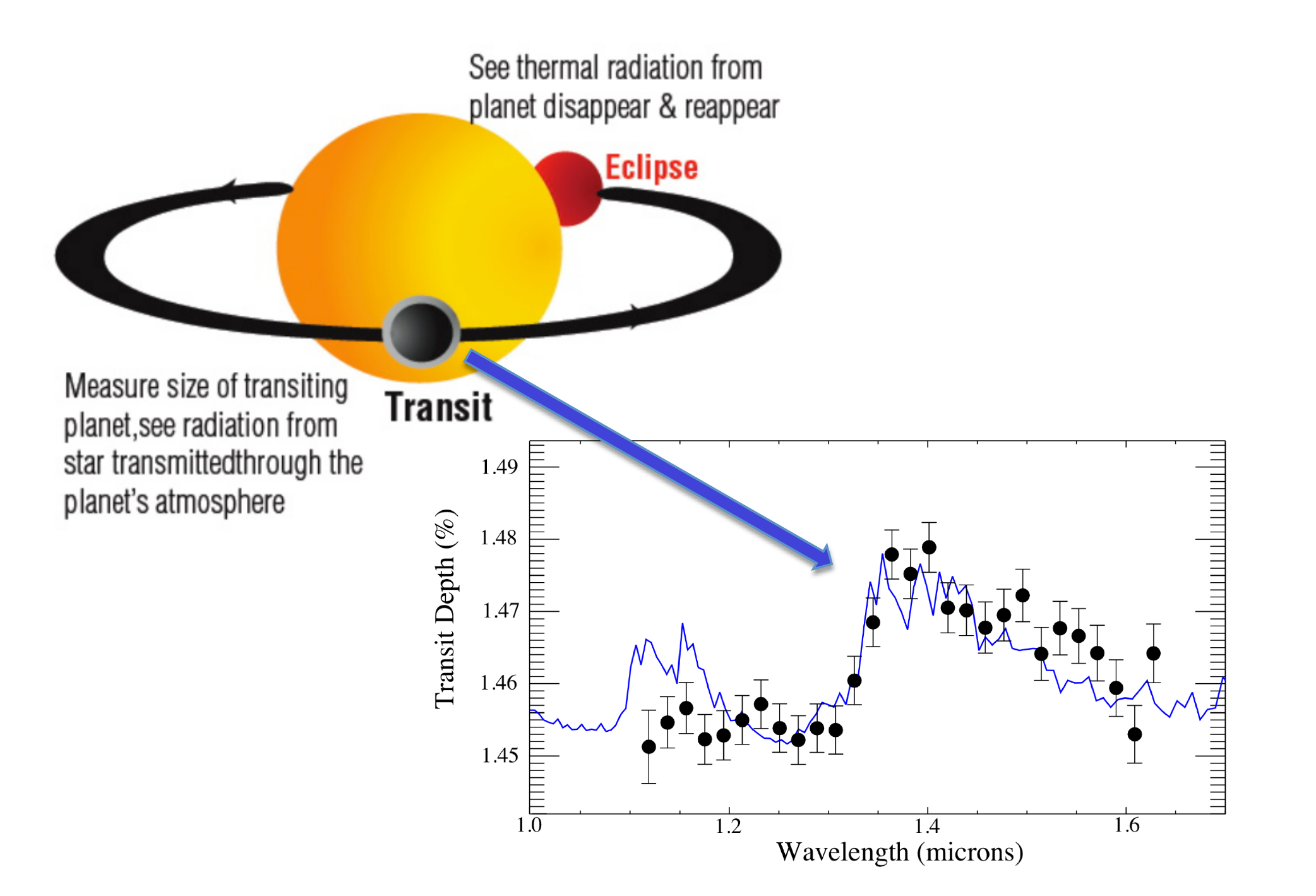}
\caption{Geometry of a transiting extrasolar planet, showing the transit when the planet passes in front of the star as seen along our line of sight, and the secondary eclipse when the planet passes behind the star as seen along our line of sight. }
\label{cartoon}
\end{figure}

\label{subsec:principles}
Figure~\ref{cartoon} shows the geometry of a transiting planet.  Transits require a planetary orbital inclination near 90-degrees. The inclination of the orbit is defined as the angle between the orbital plane and the plane of the sky \citep{smart}, so an inclination of 90-degrees places our line of sight in the orbital plane.  The terminology for transit events is inherited from eclipsing binary stars (e.g., \citealp{kalrath}), but some differences in usage are common. The time when the planet passes in front of the star is known as the transit (or, the primary eclipse for binary stars).  The time when the planet passes behind the star is known as the secondary eclipse, but is often called simply "the eclipse," or the occultation \citep{winn10}.  A transiting planet whose orbit is circular must have a secondary eclipse if it has a transit.  Planets on very eccentric orbits can have either a transit or a secondary eclipse, but not necessarily both.  Transiting planets that do not eclipse, and eclipsed planets that do not transit, are both rare.

Under the very likely assumption that exoplanetary orbital planes are aligned at random relative to our line of sight, the transit probability for a close-in planet in a circular orbit is $R_s/a$ \citep{winn10}, where $R_s$ is the stellar radius and $a$ is the radius of the orbit.  Hence, planets are more likely to transit their host stars when they have short orbital periods (small $a$ corresponds to short orbital period, from Kepler's third law).  Close-in planets experience a strong tidal force from their host star, that can circularize their orbits on a time scale that is short compared to the age of the system \citep{guillot96, jackson08}, and there is observational evidence that close-in orbits are circularized by classic tidal interactions \citep{pont11}. Transiting planets are therefore likely to have circular, or near circular orbits (albeit with some spectacular exceptions, see \citealp{laughlin09}).  In order to probe the atmosphere using secondary eclipses and (especially) phase curves (see below), we need to know the eccentricity and spatial orientation of the orbit.  We know from radial velocity and transit observations that the orbits of close-in transiting planets are approximately circular.  

There are three fundamental principles that apply when using transits to measure an exoplanet's atmosphere.  First, we cannot (so far) spatially resolve the planet from the star, and all of our transit measurements utilize the combined light of the system (star+planet). We are very far from being able to make an image of the geometry shown in Figure~\ref{cartoon}.   Second, the atmospheric signal can only be measured to the extent that it is {\it modulated in time}.  For example, as the planet passes in front of the star as seen from Earth (Figure~\ref{cartoon}), we can observe absorption features due to the planet's atmosphere appear and disappear in the combined light of the system, synchronously with the transit.  A different type of temporal modulation occurs when the wavelengths of the exoplanetary spectral features are strongly Doppler-shifted due to the orbital motion of the planet.  That Doppler shift repeats synchronously with the orbit of the planet, so that modulation occurs in both time and wavelength, and that modulation is a good way to detect the emergent spectrum of the planet (e.g., \citealp{birkby17}, see Section~\ref{subsec:convol}).  

The reason that only modulated signals can be detected bears explanation.   There are many potential sources of interference that can create spurious spectral features, and we do not understand all of those sources of non-planetary signals to the very high level of fidelity that would be required to correct for them in an absolute sense.  Instead, we are able to discriminate against the non-planetary signals because they are usually not synchronous with the transit.  

A third principle is that transit measurements are self-calibrating in terms of flux.  For example, the flux decrease at transit is used to measure the ratio of the planetary and stellar radii \citep{charbonneau00}, so in that measurement the absolute flux from the star cancels (but see \citealp{rackham18}).  Also, the emergent flux from the planet at secondary eclipse is measured relative to its host star. That relative measurement (called "contrast") is then converted to physical units using photometry of the star and a stellar model atmosphere to establish the absolute flux level.

\subsection{At What Wavelength Should We Observe?} 
\label{subsec:wavelength}

The wavelength at which we observe a transiting exoplanet determines the properties of the emitted or reflected light that we measure.  Some transiting planets orbit within $\sim 6$ stellar radii of their host star.  At that distance, they are heated to 3000K, or even hotter \citep{gaudi17}.  They are nevertheless cooler than their host star, and their contrast at secondary eclipse increases with wavelength. Recall that contrast is the {\it ratio} of the flux from the planet to that of the star, and contrast peaks at the longest observable wavelength.  The flux from the planet itself peaks at a shorter wavelength, often in the near infrared near 2\,$\mu$m, depending on the planet's temperature.  

Because the star is usually the principal source of noise, secondary eclipse measurements tend to be more favorable (in terms of observed signal-to-noise) when the contrast is maximum, i.e. in the infrared spectral region.  Planets emit their own thermal radiation because they are warm, and they also reflect stellar light.  The magnitude of thermal emission relative to reflected light increases with wavelength, for two reasons.  First, at the temperatures of transiting exoplanets, the Planck function (hence, planetary thermal emission) peaks in the infrared longward of 1\,$\mu$m.  Second, the light emitted by the star usually peaks in the optical (except for M-dwarf stars), and decreases at longer wavelengths. Hence reflected light from solar-type (FGK) stars decreases with wavelength longward of 1\,$\mu$m.  Therefore, the optical spectral region is the domain of reflected light from exoplanets, and the infrared is the domain of thermal emission from exoplanets.

\begin{figure}
\includegraphics[width=3in]{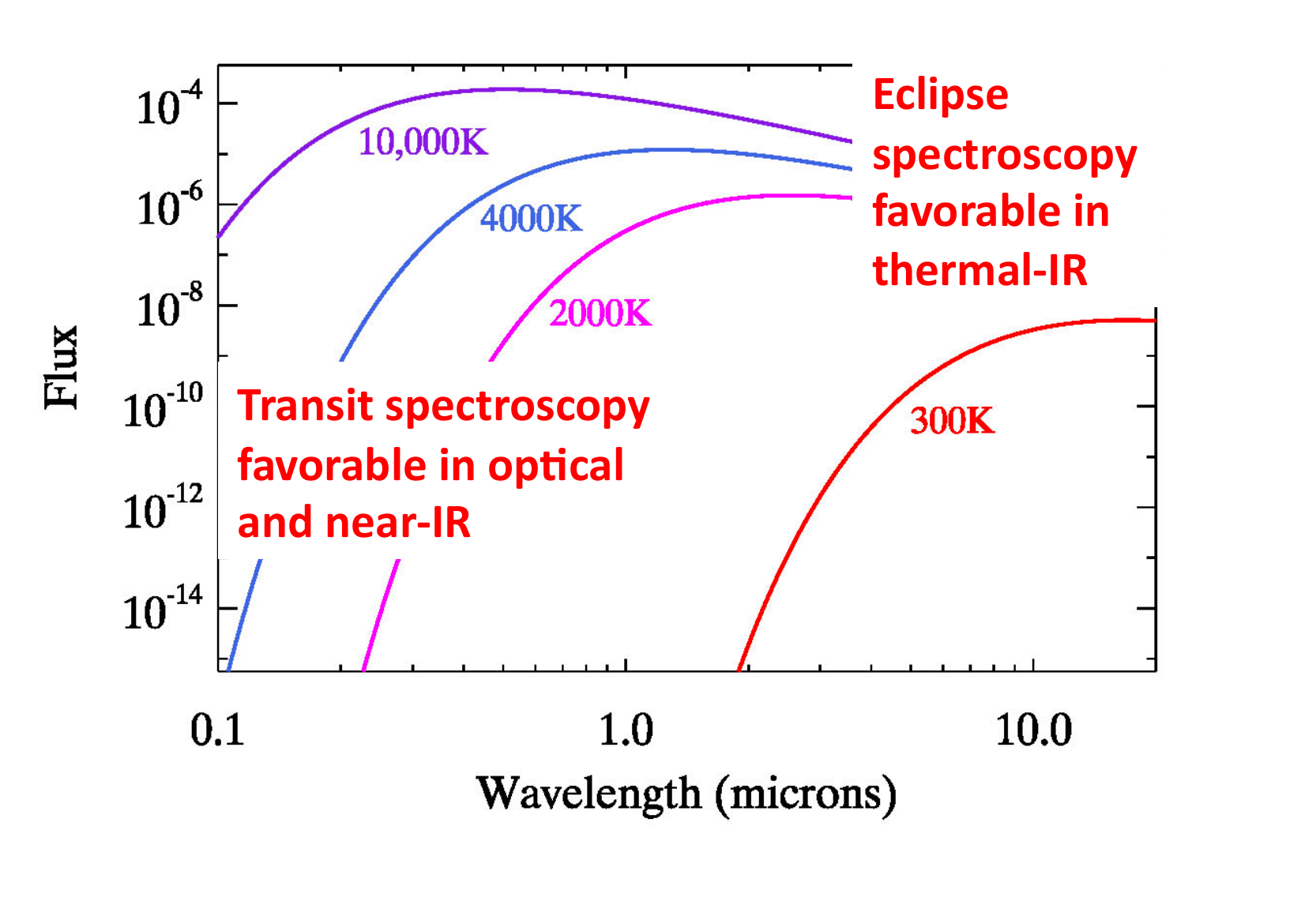}
\caption{Wavelength dependence of the Planck function for a range of temperatures than spans solar-type stars and transiting planets.  Although stellar spectra deviate from the blackbody spectra pictured here, the overall continuum level in stellar and planetary spectra are similar to blackbodies.  This illustrates that transit spectroscopy is more favorable at optical wavelengths, where the illuminating source is brightest (the star produces both signal and noise for transit spectroscopy).  The star is only a source of noise for eclipse spectroscopy, that is more favorable at the longer wavelengths where emission from the planet peaks. }
\label{fig_wavelength_range}
\end{figure}

Generally speaking, spectroscopy at transit - wherein stellar light is transmitted through the planetary atmosphere - is more favorable at optical wavelengths where the star is brightest, as illustrated in Figure~\ref{fig_wavelength_range}. Spectroscopy at transit relies on the star being the illuminating source, and (all else being equal) the signal-to-noise ratio for transit spectroscopy increases as the square root of the brightness of the star.  Transit spectroscopy at long infrared wavelengths is much more difficult, because the illuminating star is much fainter than it is at shorter wavelengths.  An exception is that cool stars such as M-dwarfs are brightest in the near-infrared ($\sim 2\,\mu$m), so transit spectroscopy of their planets is favorable at those wavelengths.

\begin{table}
\label{contrast_table}
\centering\caption{Secondary eclipse depths for thermal emission, and transit absorption signals for one scale height of absorption, in parts-per-million.  Two planets are listed: a hot Jupiter with mass of $1M_{J}$ and radius of $1.2R_{J}$ at 2000K temperature, and a 'cool' (T=500K) super-Earth with mass of $3.4M_{\oplus}$ and radius $1.5R_{\oplus}$.  The hot Jupiter transits a solar-type star, and the super-Earth transits a red dwarf star having a radius of $0.2R_{\odot}$ and a temperature of 3000K.  A solar abundance atmosphere is adopted for the hot Jupiter, and a carbon dioxide atmosphere is used for the super-Earth.  The secondary eclipse depth (in 'contrast' units, see text) depends on wavelength, and the depths for the super-Earth are essentially zero at the two shortest wavelengths (negligible thermal emission).  The transit absorption signals (also listed in parts-per-million) are for one scale height of absorption, and are independent of wavelength. }
\begin{tabular}{lllll}
Planet &  0.7\,$\mu$m  &  2.0\,$\mu$m  &  10\,$\mu$m  & Transit signal \\
\hline
\hline
Hot Jupiter & 17 & 995  &  3849  & 137 \\
Super-Earth & 0  &  0.03  & 172  & 6.3 \\
\end{tabular}
\end{table}

We now discuss how various aspects of the transit geometry are used to measure exoplanetary spectra.  Although the actual transit event was the first aspect to be discovered \citep{charbonneau00} and exploited for atmospheric measurements \citep{charbonneau02}, we here arrange our discussion in order of the approximate magnitude of the measured {\it atmospheric} signal, from largest to smallest.  A recent review of techniques used for measuring spectra of transiting exoplanets is given by \citet{kreidberg17}. Table~1 gives the typical magnitude of the atmospheric signal from a transiting Jupiter-like planet orbiting a solar-type star, and an Earth-like planet orbiting an M-dwarf star, in units of the stellar brightness.  

\subsection{Secondary Eclipses} 
\label{subsec:eclipses}

When a transiting planet disappears behind its star, as seen from our line of sight, the total brightness of the system decreases by an amount equal to the radiation emitted or reflected by the planet.  At infrared wavelengths, light {\it emitted} by the planet tends to dominate.  After some pioneering early attempts using ground-based telescopes \citep{richardson03}, the first secondary eclipses were detected using the Spitzer Space Telescope \citep{charbonneau05, deming05}, and the many contributions of Spitzer to exoplanetary science are reviewed by \citet{beichman18}.  

\begin{figure}
\includegraphics[width=3in]{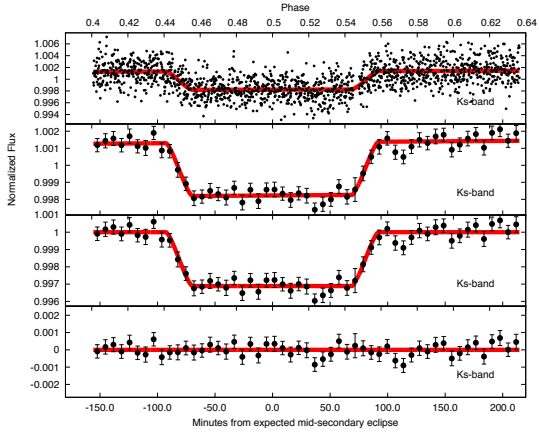}
\caption{Secondary eclipse of the hot giant exoplanet WASP-12b, measured from the ground in the K-band (2\,$\mu$m) by \citet{croll11}.  The top panel shows the measured data, and the middle panels show binned data for the same eclipse.  The red curve is a model fit, and the bottom panel shows the binned residuals after subtraction of the model from the data.}
\label{wasp12_fig}
\end{figure}

Secondary eclipse depths in the infrared can be as large as several times 0.1-percent relative to the star, which is a large signal by the standards of this field.  The highest signal-to-noise eclipses are obtained sensing the thermal emission of the planet using Spitzer \citep{knutson08, todorov12, deming15, kilpatrick17, garhart18a, garhart18b, kreidberg18}.  However, in favorable cases good eclipse measurements can be obtained using ground-based telescopes in the K-band near 2\,$\mu$m \citep{croll11, croll15, cruz16, martioli18}, or the Hubble Space Telescope at 1.4\,$\mu$m \citep{sheppard17, arcangeli18, kreidberg18}.  For the hottest planets, even optical measurements ($\lambda < $1\,$\mu$m) can sometimes detect the secondary eclipse \citep{alonso09, snellen10, lopez-morales10, fogtmann14}. Figure~\ref{wasp12_fig} shows a secondary eclipse of the very hot giant exoplanet WASP-12b, as measured at a wavelength near 2\,$\mu$m by \citet{croll11} using a ground-based telescope. Eclipses of giant planets as cool as $\sim 800$K \citep{kammer15}, or hot planets approaching the size of the Earth \citep{demory12, demory16a, tamburo18}, can be measured using Spitzer.  A summary of all secondary eclipse measurements (complete through December 2017) is given by \citet{alonso18}, and a large number of additional measurements are forthcoming from \citet{garhart18b}.

At wavelengths where the star is brightest, a significant fraction of this secondary eclipse can be due to the disappearance of stellar light {\it reflected} by the planet \citep{seager00}. 

Secondary eclipses can be observed either spectroscopically, or using photometry. When data are obtained for a secondary eclipse, the eclipse is often not immediately apparent because the data usually contain other "signals" produced by the instrumentation.  For example, spatial variation in the sensitivity of detector pixels, combined with even a small amount of pointing jitter in the telescope, can produce an effect that is much greater than a secondary eclipse.  Those signals (often called "instrumental systematics") must be removed from the data before the eclipse can be identified and interpreted \citep{stevenson12}.  That process of removing the instrumental systematics is called decorrelation, because some of the earliest methods to accomplish it relied on isolating the instrumental signal by correlating the data with some other measurable parameter such as the position of the star on the detector.  Ideally, the decorrelation process would be accomplished as part of the analysis that measures the eclipse depth (see below), but in practice it is sometimes more convenient to accomplish these as separate tasks (iteratively): first a decorrelation to remove the instrument signals, then an analysis to measure the eclipse depth.

It is appropriate to comment on the general methods used to separate secondary eclipse signals from instrumental signatures.  Bayesian methods are commonly used \citep{parviainen18}, and sophisticated parameter-search algorithms are invoked to explore the combination of instrumental and astrophysical parameters that can account for the observed data.  One question that arises is how to treat the orbital parameters of the planet in this process.  One possibility is to use priors for the orbital parameters based on transit and radial velocity (RV) observations, and allow them to vary within some error range during the data-fitting process (e.g., \citealp{martioli18}). However, we believe that using priors is often inappropriate (especially independent Gaussian priors on individual orbital parameters). Secondary eclipses are intrinsically weak compared to a transit of the entire planet (as opposed to the atmospheric signal during transit). The signal-to-noise on the eclipse depth and shape is inferior to entire-planet transit data.  Using priors when fitting eclipse data can accept a combination of orbital parameter values that would be rejected by a fit to transit data for the same planet.  The best procedure would be to perform a simultaneous fit to all available transit and eclipse data for a given planet. However, that is often not possible, or is awkward to accomplish.  We believe that the next-best alternative is to freeze the orbital parameters at the values determined by transit photometry and RV observations, fitting only for the eclipse depth and the central phase of the eclipse (that derives from the orbital eccentricity and longitude of periastron).

Once a secondary eclipse has been measured observationally, the eclipse depth must be transformed into the emergent spectrum of the planet.  In the common case where the eclipse is measured using a broad photometric bandpass, the eclipse depth in band $i$ is given as:

\begin{equation}
D_i = \frac {\int R_p^2 I_p(\lambda)R(\lambda) d{\lambda}}  {\int R_s^2 I_s(\lambda)R(\lambda) d{\lambda}}
\label{depth_equation}
\end{equation}

where $I_p(\lambda)$ is the specific intensity emitted by the planet as a function of wavelength within the bandpass, $I_s$ is the specific intensity from the star in that bandpass, and $R(\lambda)$ is the relative response of the instrumentation in that band. $R_p$ and $R_s$ are the planetary and stellar radii, respectively, and their squares appear in order to represent the solid angles that they subtend as seen from Earth.  $I_s$ is estimated using a model stellar atmosphere, based on a spectroscopically-determined stellar temperature. Note that the distance to the system does not enter because it is common to both the planet and star, and cancels when we write the above equation in terms of flux, leaving us with specific intensity.  Also, the above expression equates the stellar flux to an intensity times the disk solid angle, but the intensity varies with position on the stellar disk due to limb darkening.  For most secondary eclipse work, the intensity at disk center can be used as an acceptable approximation to the average intensity over the disk.

It is not generally possible to invert Equation~\ref{depth_equation} to transform the measurement of $D_i$ to the flux from the planet ($R_p^2I_p(\lambda)$) directly.  Instead, a model of the stellar atmosphere is adopted and the intensities emitted by that stellar model, and by a candidate model of the exoplanetary atmosphere, are integrated over the bandpass of the instrument.  The ratio of those integrals is multiplied by the ratio of the disk areas (planet divided by star, the $R^2$ are factored out of the integrals in Equation~\ref{depth_equation}), and the result is the expected eclipse depth for that particular model of the planet's emergent spectrum.  The spectrum of the planet is by far the most uncertain quantity in this analysis.  The instrumental bandpass and the parameters of the star (and the stellar model atmosphere) are well known in comparison to the spectrum of the planet. 

An especially strong observational constraint on the eccentricity of the planet's orbit can be obtained from the time of secondary eclipse \citep{charbonneau05, deming05}, combined with RV data for the host star \citep{knutson14a}.  Even an orbital eccentricity as small as $0.01$ - difficult to measure using RV data alone - can change the timing of the secondary eclipse by an hour or more, which is easily measured for many planets \citep{garhart18b}.  If we denote the orbital phase of the transit as phase $0$ or $1$, the eclipse time will be offset from phase $0.5$ by $\delta{t} = \frac{2P{e}\cos{\omega}}{\pi}$ \citep{charbonneau05}, where $P$ is the orbital period, $e$ is eccentricity, and $\omega$ is the longitude of periastron \citep{charbonneau05}. Note that the eclipse time constrains ${e}\cos{\omega}$, not $e$ alone. Note also that the previous formula is only valid in the limit of small eccentricities ($e \lessapprox 0.05$), for the full expression see \citet{pal10}.  The duration of the eclipse depends on a different combination of $e$ and $\omega$, but for most secondary eclipses the duration is difficult to measure with sufficient precision to be useful in constraining the orbital parameters.

\subsection{Phase Curves} 
\label{subsec:phasecurves}

\begin{figure*}
\includegraphics[width=6in]{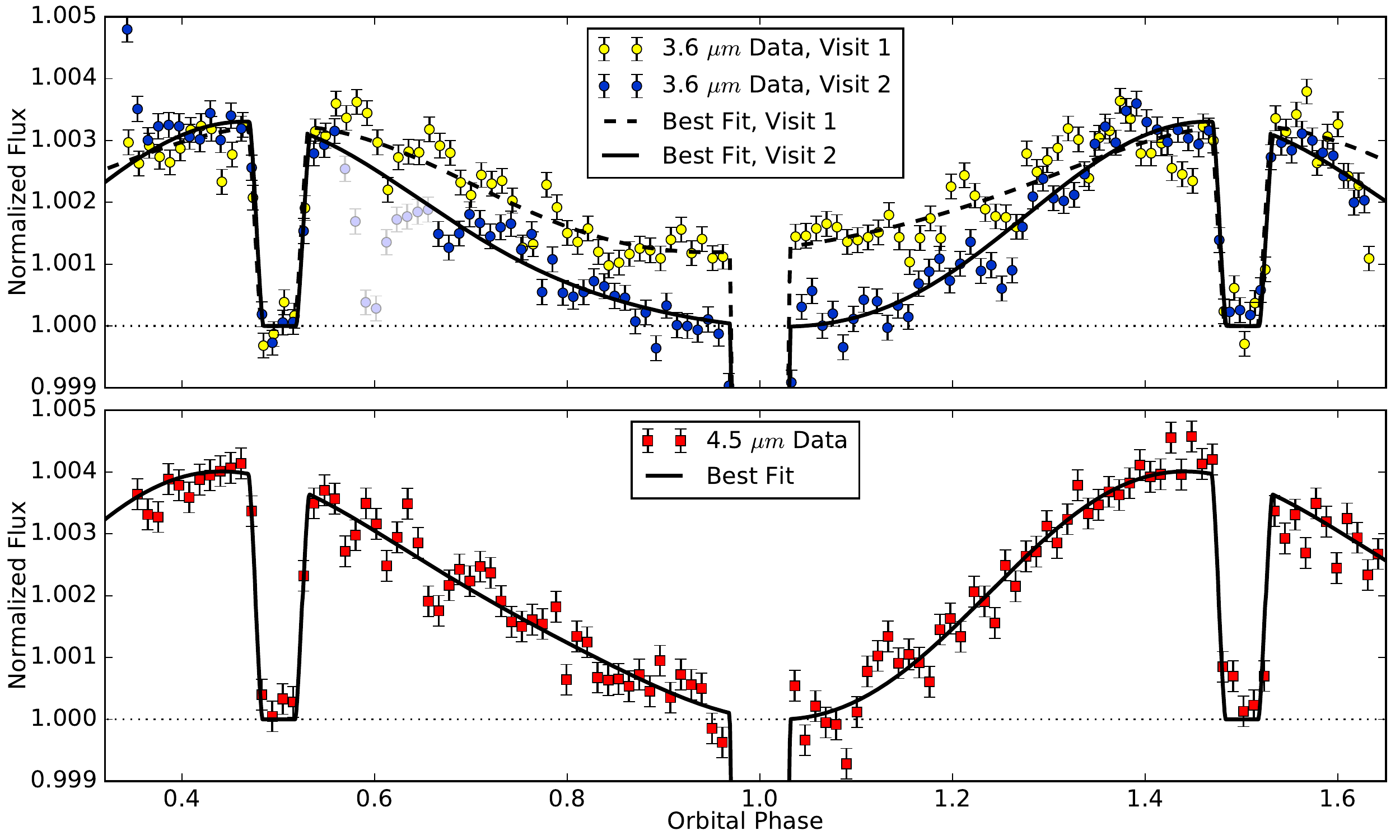}
\caption{Phase curves measured for the strongly irradiated planet WASP-43b, by \citet{stevenson17} using the 3.6- and 4.5\,$\mu$m bands of the Spitzer Space Telescope. The transit occurs at phase 1.0, and is off-scale on this plot.  Secondary eclipses are evident at orbital phases of 0.5 and 1.5. Note that the phase curve amplitude varies between two visits at 3.6\,$\mu$m, possibly because of the effect of clouds (although \citet{stevenson17} did not claim that.) }
\label{stevenson_fig}
\end{figure*}

The rotational periods of transiting planets tend to be synchronous with their orbital periods, and the planetary rotation is tidally locked to the orbit.  In the case of a 1-to-1 ratio of orbital and rotational period, we view the star-facing hemisphere of the planet just before and after secondary eclipse, and the anti-stellar hemisphere near transit.  If we observe the planet over a full orbit, we receive radiation emitted from all regions of the planet.  If the planet's spin axis is perpendicular to the orbital plane (i.e., rotational and orbital angular momenta are parallel), then we receive emergent radiation as a function of longitude on the planet, with a specific longitude dominating at any specific time.  An observation of this type over a full orbit (or longer) is called a phase curve, and an example is shown in Figure~\ref{stevenson_fig}. The more that the temperature of the planet varies spatially, the greater the amplitude of the phase curve.  To the extent that the temperature of the planet is spatially uniform, the amplitude of the phase curve would tend toward zero, except for the secondary eclipse portion.  Except for the eccentric planet HAT-P-2b \citep{lewis13}, the amplitude of observed phase curves are less than or equal to the magnitude of the secondary eclipse, although some spatial distributions of planetary temperature could in principle (albeit, unlikely) produce phase curves for non-eccentric planets whose amplitude exceeded that of the secondary eclipse. 

We can use phase curves to infer many properties of an exoplanet's atmosphere.  Following the pioneering studies by \citet{harrington06} and \citet{knutson07}, this sub-field has exploded, with phase curves measured for numerous planets using both Spitzer (e.g., \citealp{crossfield12, knutson12, lewis14, wong15, wong16, dang18, zhang18}) and Kepler (e.g., \citealp{demory13, esteves13, heng13, esteves15, angerhausen15, fogtmann14, gelino14, shporer15, jansen18}).  Although HST's near-Earth orbit usually does not allow continuous observing, HST data have also yielded valuable phase curves \citep{stevenson14, kreidberg18}.  Because phase curves do not spatially resolve the planet, the inherent resolution in longitude is limited to about 5 independent values per planetary circumference \citep{cowan08, cowan17}.  Phase curves are nevertheless a powerful technique to learn about the energy balance, atmospheric dynamics (i.e., winds), and cloud formation in the exoplanetary atmosphere.  

Phase curves that are measured at infrared wavelengths are sensitive to planetary temperature via thermal emission. Phase curves measured at optical wavelengths are sensitive to reflected light, but thermal emission can also contribute at optical wavelengths when the planet is very hot. The first well sampled infrared phase curve \citep{knutson07} revealed an eastward offset between the hottest measured longitude and the most strongly irradiated longitude at the sub-stellar point.  Circulation models for tidally locked hot Jupiters \citep{showman02,cho03} predicted eastward winds\footnote{Eastward = toward the east = westerly} driven by the re-distribution of stellar heat.  Those winds advect the hottest spot, i.e., the winds move heat by fluid motion.  Competing with advection is radiative cooling, so that the hottest gas cools by radiation as it moves. Consequently, the hottest regions of the planet are eastward of the sub-stellar point, to a degree that is determined by the balance between advection and radiative cooling \citep{showman11, perez-becker13, komacek16}. That creates an offset in the peak of the phase curve, because the hottest part of the planet is aligned with the direction to Earth slightly prior to phase 0.5.  A peak that occurs prior to secondary eclipse has been observed consistently for multiple planets, but not all \citep{dang18}.  Also, because the Earth-facing hemisphere of the planet is not spatially uniform, the center time of secondary eclipse \citep{williams06} is slightly offset from phase 0.5, but the secondary eclipse offset ($\sim 100$ seconds) has not yet been definitively observed (but see \citealp{agol10}).

\begin{figure}
\includegraphics[width=3in]{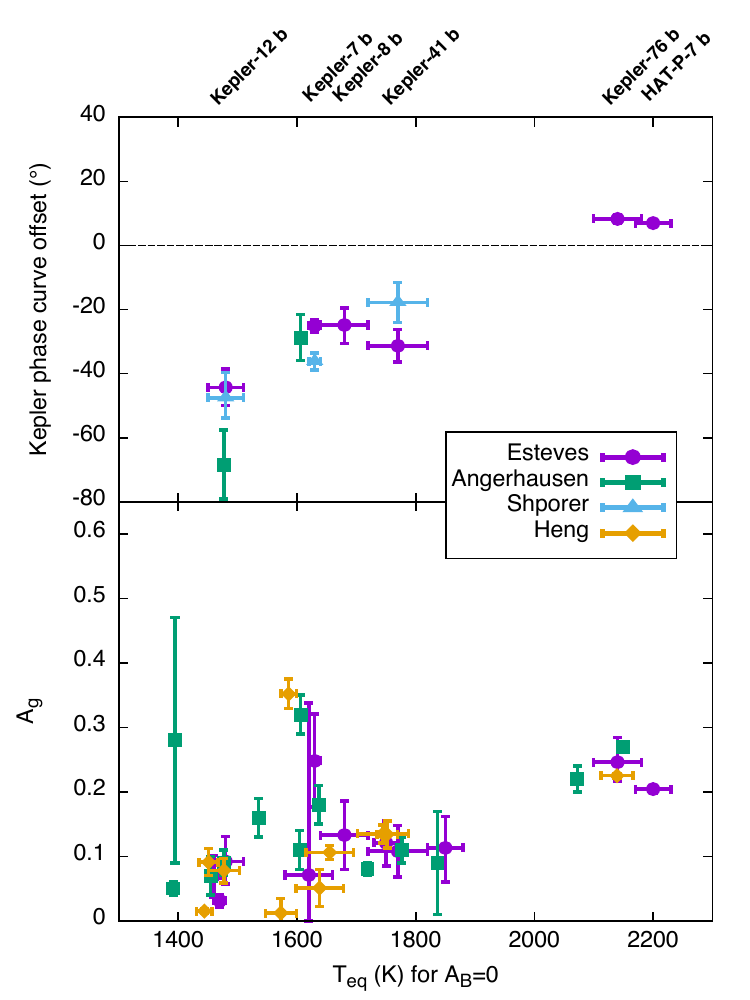}
\caption{Phase shifts measured from optical-wavelength phase curves in Kepler data measured by \citet{esteves15}, \citet{angerhausen15}, \citet{shporer15}, and \citet{heng13}.  The phase shift is plotted versus planetary equilibrium temperature, assuming zero albedo.  This Figure is taken from \citet{parmentier16}.  The peak of the phase curve in optical light depends on the dominant longitude where clouds are formed. \citet{parmentier16} find that the sub-stellar point is usually too hot for clouds to form, so clouds tend to form in the cooler regions west of the sub-stellar point (negative phase shifts).  The phase shift varies with the equilibrium temperature of the planet because clouds of different composition condense at different temperatures, requiring different longitudes.  For the hottest planets, thermal emission begins to be important and the phase shift is determined by what longitudes are hottest.  That causes a phase shift to the east, determined by the balance between winds and the time scale for radiative cooling. }
\label{parmentier_fig}
\end{figure}

Optical phase curves can reveal the nature of cloud formation \citep{oreshenko16, parmentier16}.  Reflected light from clouds will tend to cause a peak at a particular time in the phase curve, because time corresponds to longitude on the planet, and hence to temperature.  Clouds of various composition will condense at different temperatures, so they form and reflect star light at different distances from the sub-stellar point. \citet{oreshenko16} found that the measurement uncertainties were too large to distinguish various compositions for the clouds of HAT-P-7b, but \citet{parmentier16} were more optimistic in discriminating between different cloud compositions. Moreover, the temperature distribution on the planet changes as a function of the planet's equilibrium temperature. Consequently, the peak in an optical phase curve is a function of the equilibrium temperature of the planet, shown in Figure~\ref{parmentier_fig}, from \citet{parmentier16}.  

Clouds can also affect phase curves in thermal emission. A cloud at high altitude can block radiation that would otherwise emerge from deeper, hot layers.  Consequently, the interpretation of phase curves can be complex, because optical phase curves exhibit reflected light, but can a contain contribution from thermal emission.  Infrared phase curves exhibit thermal emission, but can also be affected by the opacity of the same clouds that reflect optical light.

In addition to the longitudinal distribution of heat and reflected light from clouds, phase curves are also sensitive to other effects, such as tidal distortion of the star (induced by the planet) and relativistic beaming \citep{loeb03, mazeh10, barclay12, esteves13}. Tidal distortion can cause both the star and planet to subtend a larger solid angle (hence, be brighter) when the planet is approximately at the greatest elongation from the star as seen along our line of sight, thus having a different dependence on orbital phase than do reflected or emitted light from the planet. Relativistic beaming causes the planet's emergent or reflected light to increase or decrease, depending on the planet's radial velocity.  Relativistic beaming affects both the total amount of radiation, as well as the fraction within a given observational bandpass \citep{barclay12}.  Although neither relativistic beaming nor tidal distortion effects give us direct information concerning the exoplanetary atmosphere, it is possible to separate them from the effect of the longitudinal heat distribution and reflected light.  A review of optical phase curve effects is given by \citet{shporer17}. 

\subsection{Transits} 
\label{subsec:transits}

Absorption of stellar light that passes through the atmospheric annulus of a planet during transit \citep{seager_sasselov, brown01, hubbard01} was the method used by \citet{charbonneau02} to make the first detection of an exoplanet's atmosphere. Transit measurements are relative in the sense that the flux of the system (star + planet) during transit is compared to the out-of-transit flux. At wavelengths where the atomic or molecular constituents of the exoplanetary atmosphere absorb strongly, the atmospheric annulus (see Figure~\ref{cartoon}) can be opaque over several scale heights. In all cases, the depth of the transit at a given wavelength is defined as:

\begin{equation}
D(\lambda) = (F_{out} - F_{in})/F_{out},
\end{equation}

where $F_{out}$ is the measured flux of the system (star + planet) out of transit and $F_{in}$ is the flux during transit (and these fluxes depend on wavelength). Since $F_{in} < F_{out}$, the definition above conveniently insures that the transit depth is positive, whereas physically the transit is negative, i.e. represents a decrease in the stellar flux.  Flux is proportional to intensity $I$, times the solid angle subtended by the portion of the stellar disk that is not covered by the planet.  Since the star and planet are (to an excellent approximation) at the same distance from us, we can use cross-sectional areas in lieu of solid angles.  In that case, we have:

\begin{equation}
F_{out}=I{\pi}R_{s}^2,
\end{equation}

which simply says that the flux out of transit is the flux from star alone (because emission from the planet can be neglected in most transit applications).  In transit, we have:

\begin{equation}
F_{in}=I({\pi}R_{s}^2-{\pi}R_{p}^2),
\end{equation}

which accounts for the portion of the stellar disk blocked by the planet.  The transit spectrum is the {\it variation} of $D(\lambda)$ with wavelength, assuming it is large enough to be measurable in the presence of noise.  To estimate the magnitude of that variation, we (temporarily) make two approximations.  First, we postulate that strong absorption causes the atmospheric annulus to be opaque over one scale height, and transparent over other ranges of height. The scale height of the atmosphere is given as:

\begin{equation}
H = kT/{\mu}mg,
\end{equation}

where $T$ is the temperature of the atmosphere, $k$ is Boltzmann's constant, ${\mu}m$ is the mean weight of particles comprising the atmosphere, and $g$ is the acceleration due to gravity. If we also make the approximation that the stellar disk has a uniform intensity $I$ (i.e., we ignore limb darkening), then the intensity cancels in the expression for the depth of the transit.  The transit depth at wavelength $\lambda$ becomes:

\begin{equation}
D(\lambda) = ( {\pi}R_{s}^2 - ( {\pi}R_s^2 - {\pi}(R_{p}+H)^2 ))/ {\pi}R_{s}^2, 
\end{equation}

where $R_s$ and $R_p$ are the radii of the star and planet, respectively.  Expanding the above relation, we find:

\begin{equation}
D(\lambda) = (R_{p}/R_{s})^2 + 2R_{p}H/R_{s}^2,
\end{equation}

where we have neglected a term involving $H^2$ because $H \ll R_{p}$.  The first term above is the contribution to the transit depth by the disk of the planet without the atmosphere, and the second term expresses the atmospheric contribution, that is wavelength-dependent.  At this stage we can relax our temporary assumption that the atmosphere is opaque over exactly one scale height.  Instead, we write that $A_{\lambda}$ scale heights are opaque at wavelength $\lambda$ (following \citealp{stevenson16a}, also see \citealp{iyer16} and \citealp{fu17}).  Since we are interested in the wavelength variation of $D(\lambda)$, we drop the $(R_{p}/R_{s})^2$ term, and write: 

\begin{equation}
D^{'}(\lambda) = 2R_{p}H A_{\lambda}/R_{s}^2
\end{equation}

\begin{figure}
\includegraphics[width=3in]{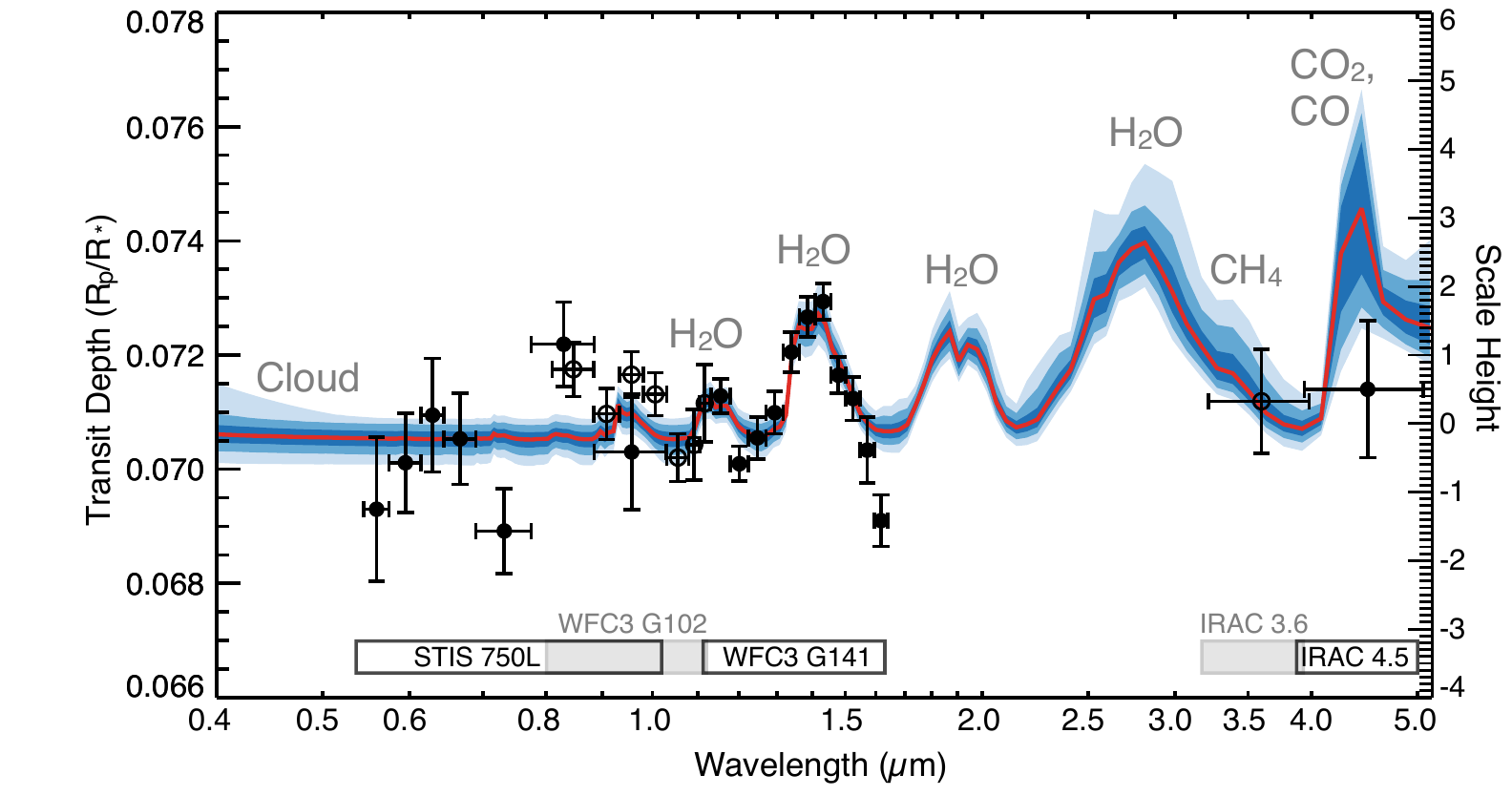}
\caption{Transit spectrum of HAT-P-26b from \citet{wakeford17} using the Hubble Space Telescope. The peak absorption near 1.4\,$\mu$m is due to water vapor. }
\label{hat26}
\end{figure}

Recalling that our sign convention makes the transit depth positive, wavelengths of strong absorption in the atmosphere have larger transit depths than wavelengths of weak absorption.  This implies that plots of transit absorption spectra will be "upside-down" (as per Figure~\ref{cartoon}) compared to normal astronomical spectra.  An example of water vapor absorption in the transit spectrum of the Neptune-mass planet HAT-P-26b, as observed by \citet{wakeford17} using the Hubble Space Telescope, is shown in Figure~\ref{hat26}.

\subsection{Eclipse Mapping} 
\label{subsec:mapping}

\begin{figure*}
\includegraphics[width=6in]{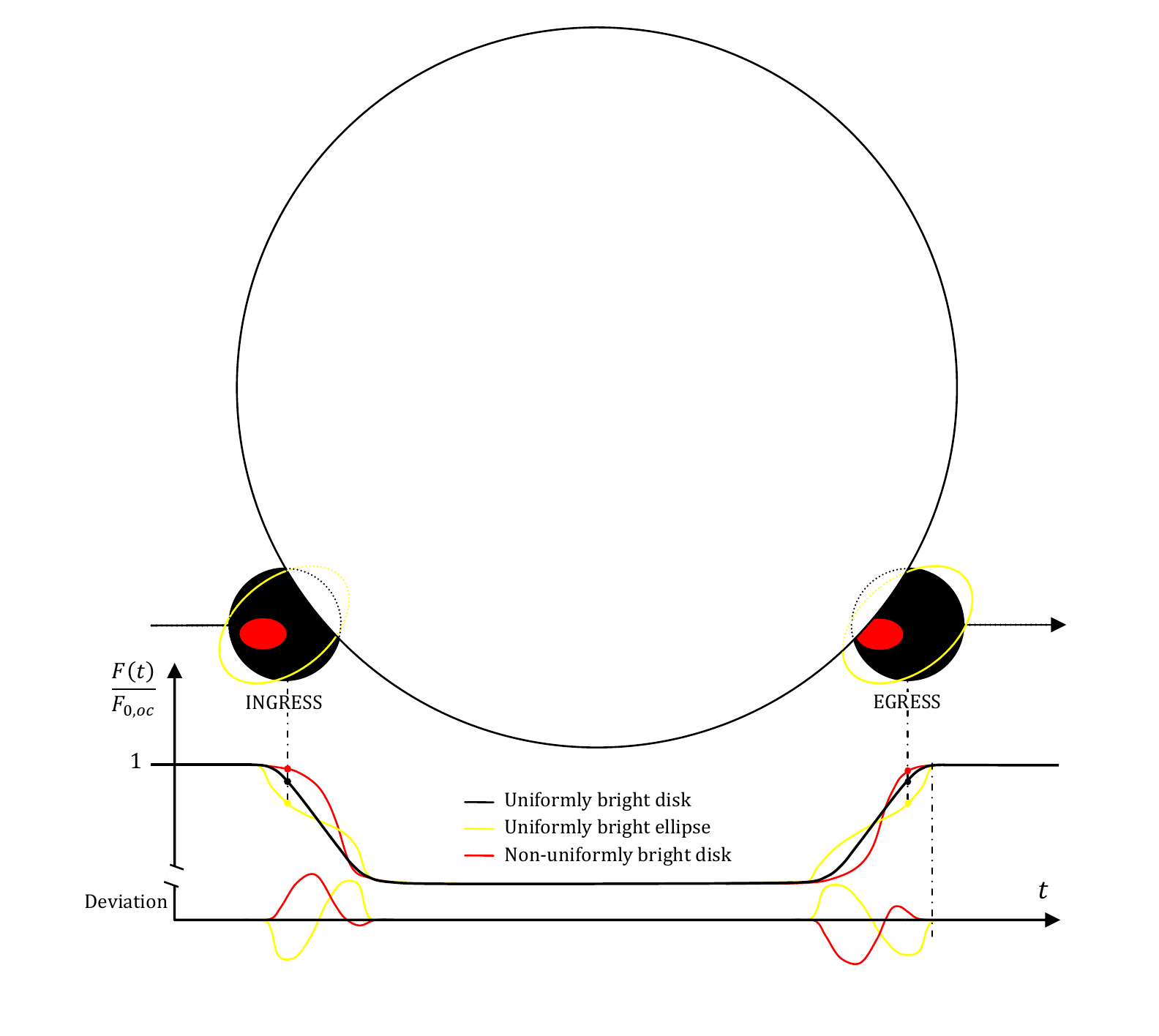}
\caption{Principle of eclipse mapping, from \citet{dewit12}. The shape of the eclipse curve at ingress and egress can be inverted to infer the presence of hot and cold regions on the disk of the planet. In this case, the hot region is represented by the red spot on the disk, and it changes the ingress/egress light curves relative to the case of a uniform disk.}
\label{dewit_fig}
\end{figure*}

When one astronomical body is eclipsed by another one, the possibility exists to map the spatial structure of the first body.  This eclipse mapping techniqe is not original to exoplanets, it has been used in astrophysics (e.g., \citealp{taylor66}), and for bodies in our Solar System for many decades (e.g., \citealp{vermilion74}).  The scale height in the atmosphere of a solar-type (FGK) star is much less than the radius of a giant planet (or even a super-Earth).  Consequently, as a transiting planet is eclipsed by its star, the edge of the star is effectively a "knife edge" that scans the disk of the planet.  Given a sufficiently high signal-to-noise, the brightness distribution of the disk of the planet can be inferred by inverting the profile at ingress and egress (see Figure~\ref{dewit_fig}).  Eclipse mapping of exoplanets can in principle work using either reflected light, or thermal emission.  Signal-to-noise is usually higher for thermal emission, so most attempts have utilized infrared data from Spitzer.  \citet{deming06} analysed the ingress and egress profiles obtained during the first high signal-to-noise secondary eclipse measurement: HD\,189733b observed using Spitzer at 16\,$\mu$m.  They concluded that the dominant effect during ingress and egress is the {\it shape} of the planet, and with only one eclipse they were able to conclude that the planet is approximately round, but they could not meaningfully constrain the brightness distribution on the disk.

The principles of eclipse mapping using thermal emission were elucidated by \citet{rauscher07}, who concluded that the photometric accuracy of Spitzer was sufficient to distinguish among parametric models of the planetary emission, using multiple eclipses.  In this context, parametric models means that the brightness structure of the planetary disk is assumed to correspond to some analytic expression. That effectively reduces the number of unknowns that must be determined by the observations, and allows simple eclipse maps to be obtained using modest signal-to-noise. They further concluded that JWST will be able to extend eclipse mapping to non-parametric models (no prior assumptions concerning the planet's disk emission). 

\begin{figure*}
\includegraphics[width=6in]{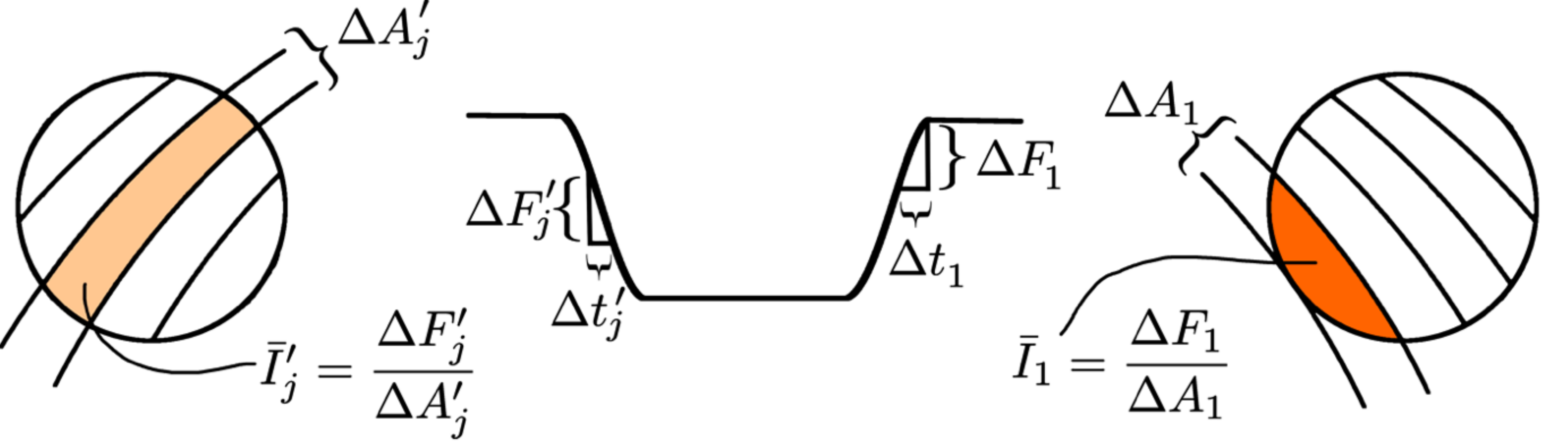}
\caption{Constructing an eclipse map having resolution in both latitude and longitude, from \citet{majeau12}. The limb of the star scans the disk of the planet at an oblique angle, and the combination of ingress and egress data yields a 2-D map of the exoplanetary disk.}
\label{majeau_fig}
\end{figure*}

In the situation where the orbit of the planet is in the same plane as its rotational equator, the limb of the star would be parallel to meridians of longitude at both ingress and egress.  That would allow only a longitudinal map of the planet to be constructed, with no information in latitude.  Fortunately, in the general case the limb of the star scans the planet at an oblique angle, and the combination of ingress and egress data can be used to construct maps with resolution in both longitude and latitude, as illustrated in Figure~\ref{majeau_fig} (from \citealp{majeau12}).

Several studies have used Spitzer data to construct spatial maps of hot Jupiters.  \citet{agol10} observed multiple 8\,$\mu$m eclipses and transits of HD\,189733b (the most easily-characterized hot Jupiter).  Among other aspects they studied the possible temporal variability of day side emission, and they mapped the spatial offset of the hottest spot, but they did not attempt to invert the data to construct spatial maps of the disk.  However, \citet{majeau12} and \citet{dewit12} and used the same data to construct a map of the disk.  They confirmed that the hottest region of the atmosphere is advected by prograde zonal winds\footnote{prograde=in the same direction as planetary rotation}, in agreement with the Spitzer phase curve \citep{knutson07}, and they also verified that this hot spot was confined to near-equatorial latitudes.  However, given the modest aperture of Spitzer, these studies developed the methodology of eclipse mapping, but they did not achieve the full potential of the technique, and significant further advances are expected using JWST.  With the larger photon flux collected by JWST, as well as its spectroscopic capability, {\it three}-dimensional mapping of exoplanetary atmospheres will by possible by exploiting the different depths of formation for radiation emitted at different wavelengths.  Eventually, it may be possible to map the velocity structure on the disk of the planet due to planetary rotation and zonal winds \citep{nikolov15}.

\subsection{Convolution with Spectral Templates} \label{subsec:convol}

\begin{figure}
\includegraphics[width=3.5in]{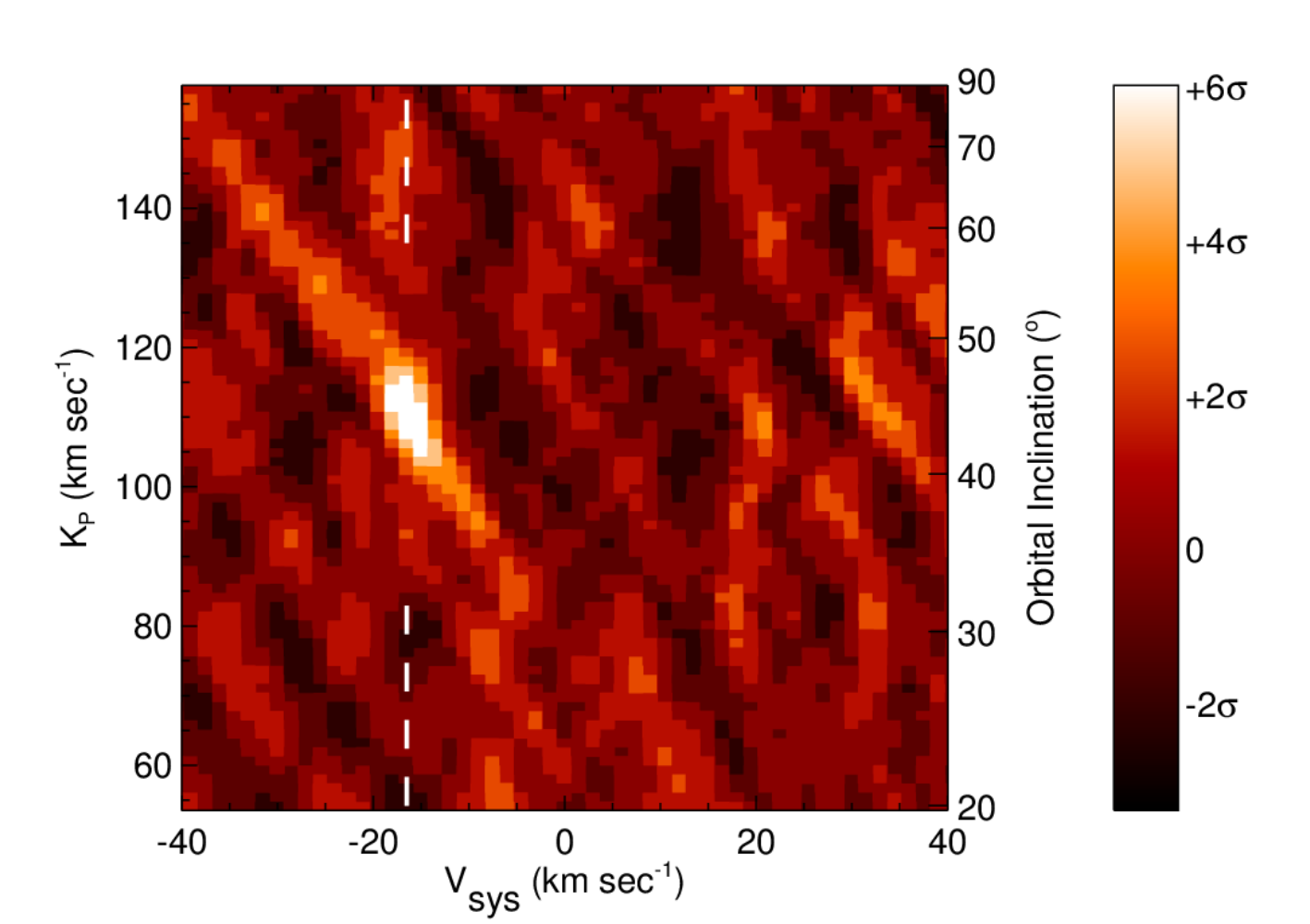}
\caption{Example of a cross-correlation detection of molecular absorption in an exoplanetary atmosphere. This detection is for carbon monoxide, by \citet{brogi12}. The brightest spot is the peak of the cross-correlation in units of standard deviations, and the sign is arranged so that the positive peak in the cross-correlation measures absorption in the emergent spectrum of the planet.  The cross-correlation is displayed as a function of the system's center-of-mass velocity on the X-axis, and the line-of-sight velocity amplitude of the planet on the Y-axis.  Because this planet does not transit, its line-of-sight velocity amplitude measures the inclination of its orbit to the plane of the sky, which is about 44 degrees.}
\label{brogi_fig}
\end{figure}

Modern quantum mechanical modeling of molecules such as water vapor \citep{barber06, polyansky18} is sufficiently accurate that calculated high resolution spectra of exoplanets should agree approximately with their actual emergent (or reflected) spectra, at least in terms of the positions and approximate relative strengths of the spectral lines.  This motivates a method to detect the molecular spectrum of a given molecule, even when individual spectral lines are beneath the threshold of detectability.  The observational data used in this method comprise a set of high resolution spectra of the star+planet system, obtained over a wide range of planetary orbital phases (hence, wide range of planetary Doppler shifts).  It can be helpful to the method if the temporal average of the spectra is subtracted from each individual spectrum, to form 'difference spectra', because the principle is to search for temporal {\it changes} in the spectrum of the star+planet system caused by the changing Doppler shift of the planet as it orbits the star.  The method thereby exploits the large orbital velocity of close-in planets, and it does not require that the planet transit, only that it has a time-variable radial velocity whose phase is known (the amplitude can be determined by this method).  The method works best when the change in radial velocity of the planet over multiple observations exceeds the width of lines in the stellar and telluric spectra.  A theoretical template spectrum is calculated, based on a model of the exoplanetary atmosphere. For each observed difference spectrum, the calculated template is Doppler-shifted to the hypothetical rest frame of the exoplanet, multiplied times the observed difference spectrum of the (star+planet) system, and integrated over wavelength.  This process produces a cross-correlation between the template spectrum and the observed difference spectra. The cross-correlation function is the dependent variable, and the independent variable can be the geocentric radial velocity of the planetary system, or (for non-transiting planets), the amplitude of the planet's radial velocity.  A significant peak in the cross-correlation function indicates a match between the template spectrum and the spectrum of the planet at that value of the independent variable, and the amplitude of that peak is a function of the degree to which the actual exoplanetary spectrum matches the template spectrum. An example of a cross-correlation peak for the non-transiting planet Tau Bootis\,b is shown in Figure~\ref{brogi_fig}.

If the amplitude of the cross-correlation is carefully interpreted, we can learn about the conditions in the exoplanetary atmosphere. In principle, cross-correlation can inform us of exoplanetary abundances, and also winds and rotation \citep{snellen10, louden15, brogi16}.  This method has been called Doppler deconvolution by \citet{deming_seager17}, and simply cross-correlation by others. Although the cross-correlation is a mathematical convolution (not formally {\it de}convolution), the net result is to exploit the changing Doppler shift of the exoplanetary spectrum to go back (i.e., deconvolve) to the spectrum of the planet.

Early attempts at cross-correlation detections of exoplanetary atmospheres \citep{wiedemann01, barnes07} were unsuccessful due to limited spectral resolving power and wavelength coverage, and lack of accurate templates.  The first successful detection was by \citet{snellen10}, and much subsequent work \citep{brogi12, brogi14, lockwood14, birkby17, hoeijmakers18} has established the viability of this technique.  It will become even more important with coupled to the large light-gathering power of very large ground-based telescopes (Section~\ref{subsec:elts}).   

\section{Photometry versus Spectroscopy} \label{sec:photspec}

Observational techniques used to characterize exoplanetary atmospheres include photometry and spectroscopy, and both techniques apply over a large range in wavelength (ultraviolet, optical, infrared).  Photometry collects radiation using a broad wavelength range (bandpass), whereas spectroscopy divides radiation into a series of narrow wavelength channels. The concept of spectral resolving power (${\lambda}/\Delta{\lambda}$) applies to both techniques, where $\Delta{\lambda}$ is the width of the most narrow wavelength channel that the instrument can measure (not to be confused with the {\it sampling} in wavelength, that is usually finer in scale than the resolution).  Photometric spectral resolving power is low, typically ${\lambda}/\Delta{\lambda} \sim 3$.  Higher spectral resolving power is always desirable; spectral line structure from exoplanetary atmospheres is often not resolved unless  ${\lambda}/\Delta{\lambda} \ge 10^{5}$.  However, the low photon fluxes available from exoplanets make high spectral resolution impractical, so photometry and low-resolution spectroscopy are the dominant techniques used to date.  As generally applied to exoplanetary atmospheres, photometry is easy to measure and hard to interpret, whereas spectroscopy is hard to measure but easy to interpret.  We explain that statement as follows.

Secondary eclipses are a prime example wherein exoplanetary atmospheres are measured using photometry.  A broad wavelength band is used to collect ample light from the star and planet, and the signals from them are separated in time via the eclipse event.  Students and new workers in the field are cautioned that {\it all} exoplanet measurements are difficult, but secondary eclipse photometry is less difficult than other techniques. Photometry collects a lot of light, and uses relatively few detector pixels. Using only a few pixels makes it easier to define the systematic errors produced by the detector (fewer unknowns). Consequently, the photon-limited precision can be excellent, and complications from the detector are not too daunting when only a minimal number of pixels are involved.  Consequently, photometry of exoplanets is relatively easy to measure.  Nevertheless, the measurements usually contain other signals that are larger than the exoplanet signal.  For example, using Spitzer a dominant signal is due to the spatial non-uniformity of the detector combined with jitter in the position of the image.  This signal is often defined by correlating the measured flux with the position of the stellar image on the detector \citep{stevenson12}. Using that correlation, the so-called intra-pixel effect can be removed, leaving the exoplanetary eclipse signal. 

Thermal emission from dozens of exoplanets have been measured using photometry \citep{alonso18}, but the measurements can be difficult to interpret. Early work in the field boldly interpreted eclipse depths at multiple wavelengths in terms of the planetary spectra, often leading to startling conclusions concerning the physical nature of the exoplanetary atmospheres \citep{madhu11}.  Stimulated by \citet{hansen14}, there is now a general consensus that inferring exoplanetary spectra from photometry is problematic.  The low spectral resolution that characterizes photometry dilutes (and thereby weakens) the strength of absorption or emission features, and makes them more difficult to detect above the noise. It also obscures their spectral structure.  For example, molecular spectra often show increased absorption at a band head where individual vibration-rotation transitions are concentrated in wavelength.  Seeing that type of characteristic structure can make the detection of a molecule unequivocal, but that is only possible when the band structure is spectrally resolved, and photometry doesn't allow that.

Spectroscopy of exoplanets is easy to interpret, provided that adequate signal-to-noise is obtained, because the characteristic shape of molecular bands can be seen in spectra.  Even strong atomic lines have characteristic structure that spectroscopy can reveal.  For example, the D-lines of sodium have strong line cores, flanked at both longer and shorter wavelengths by weaker pressure-broadened wings.  That shape makes the detection unequivocal, if sufficient signal-to-noise is obtained.  But spectroscopic measurements can be difficult because the dispersion in wavelength means that, compared to photometry, fewer photons are available at each wavelength.  Moreover, different detector pixels are (usually) used at each wavelength, and relating the properties of many different detector pixels (i.e., correcting for intrumental effects) can be more difficult than for photometry.

\section{Ground-based versus Space-borne Measurements} \label{sec:groundspace}

Many advances in the sensing of exoplanetary atmospheres have come from space-borne measurements, including the first detection of an exoplanet's atmosphere \citep{charbonneau02}, and the first detection of infrared radiation emitted by exoplanets \citep{charbonneau05, deming05}.  Exoplanetary astronomy from space (e.g., using the WFC3 instrument on HST) continues to be a very active sub-field \citep{deming13, wakeford13, mandell13, mccullough14, sing16, knutson14b, kreidberg14, wakeford17, nikolov18}. Nevertheless, ground-based astronomy contributes also, and it is appropriate to highlight what we deem to be the potential of ground-based versus space-borne measurements.

Space-borne measurements have the obvious advantage that they can be pursued at wavelengths where the telluric atmosphere is opaque (e.g., the vacuum ultraviolet).  But space-borne measurements have other advantages.  Even at nominally transparent wavelengths, telluric absorption can be variable (at wavelengths where water vapor absorbs even weakly, and at blue wavelengths where aerosols have strong scattering opacity).  Space-borne measurements eliminate that problem, and the space environment tends to be free from sources of thermal and mechanical stress that can cause irreproducible fluctuations in instrumental signals (e.g., mechanical flexure in instruments at Cassegrain foci of ground-based telescopes). 

We expect that the most important near-term results on exoplanetary atmospheres will come from space-borne facilities such as JWST (\citealp{greene16}, see Sec.~\ref{subsec:JWST}).  However, there are excellent prospects for ground-based facilities such as the Extremely Large Telescopes (ELTs, Sec.~\ref{subsec:elts}) to be dominant on a $\sim$\,decades time scale. We argue this potential dominance for several reasons.  First, the ELTs will collect many more photons than any contemporary space-borne facility.  Second, there are indications that very high precision measurements can be obtained from the ground, if instruments are appropriately designed to produce the high dynamic range that transiting exoplanet measurements require.  We note that \citet{johnson09} used a specially designed camera to achieve a photometric precision of 470 parts-per-million in a 1.3-minute exposure time, and that level of precision is competitive with space-borne measurements.  Geometric differences in the light path through the telluric atmosphere are less than the diameter of a large ground-based telescope for an exoplanet star and a comparison star lying within a few arc-minutes on the sky. Consequently, telluric fluctuations should be common to both the exoplanet system and comparison star, and differential telluric atmospheric corrections could in principle be very effective.  Corrections using comparison stars are most effective when their spectral energy distribution is similar to that of the target star, or when stellar light is dispersed in differential spectrophotometry \citep{bean10}. 

High spectral resolution studies of exoplanetary atmospheres \citep {hoeijmakers18, birkby17} are greatly facilitated by large ground-based telescopes.  Not only do large aperture telescopes collect the ample photon fluxes that high resolution spectra require, but systematic effects in the instrumentation are easier to correct when dealing with high resolution spectra.  With high resolution spectra, the exoplanetary signal is present over many different wavelengths, and those many wavelengths fall onto many different pixels of the detector.  Consequently, the exoplanet signal is not dependent on the properites of only a few pixels, but can be averaged over many pixels.  Moreover, the signal can essentially be high-pass filtered in wavelength, thereby discriminating against instrumental effects that often occur over spans of many pixels on the detector.   For example, it is common to seek the exoplanet signal by convolving the observed spectrum with an exoplanet template spectrum (see  Section~\ref{subsec:convol}). The template spectrum has fine-scale structure in wavelength due to the modeled emergent spectrum of the planet.  A convolution in wavelength space is equivalent to a multiplication in Fourier space \citep{bracewell}, and the template spectrum has much power at high spatial frequencies (i.e., strong variations occur over a span of a few detector pixels).  Hence the convolution process is similar in effect to a high-pass filter, and it helps to discriminate against instrumental systematic effects.

\section{Interpreting the Measurements} \label{sec:interpreting}

\subsection{Individual Planets}

Phase curves provide the richest opportunities for interpreting the measurements of a single planet in terms of the properties of its atmosphere.  Whereas secondary eclipses probe the day side of the planet, phase curves probe all longitudes.  In contrast with transit spectroscopy, phase curves and secondary eclipses are sensitive to the emergent spectrum of the planet (i.e., radiation that the planet emits or reflects), and that spectrum can exhibit features in either absorption or emission.  If the atmosphere is in local thermodynamic equilibrium (LTE, source function equal to the Planck funtion, see \citealp{rybicki}){\footnote{Other {\it local} conditions for LTE are that atomic and molecular level populations have Boltzmann values, ionization states have Saha values, and particle velocities are Maxwellian}, then absorption features are produced when the atmospheric temperature is declining as optical depth decreases, i.e. as height increases.  Spectral emission features are produced when temperature rises as optical depth declines, i.e. temperature rises with increasing height.  

\subsubsection{Interpreting transit Spectra}

Emergent exoplanetary spectra can exhibit either absorption or emission, but transit spectra will always exhibit absorption, except under very unusual circumstances. Absorption occurs because of the properties of radiative transfer. When stellar radiation interacts with atoms and molecules in the planet's atmosphere, it is absorbed and induces transitions between quantum states in the absorbing atom/molecule.  If $I_s$ is the intensity of stellar radiation (in photons cm$^{-2}$ sec$^{-1}$ sr$^{-1}$ Hz$^{-1}$), then the change in intensity \citet{rybicki} is given as:

\begin{equation}
{\delta}I_s = L  {I_s} (B_{ul}{N_u} - B_{lu}{N_l})  + L A_{ul} N_u
\end{equation}

where $L$ is the path length through the exoplanetary atmosphere, $N_l$ and $N_u$ are the number densities in the lower and upper quantum states respectively (for simplicity, assumed to be constant along the radiation path). $B_{lu}$ and $B_{ul}$ are the Einstein B-coefficients for absorption from the lower to upper state, and stimulated emission from the upper to lower state respectively. $A_{ul}$ is the Einstein-A coefficient for spontaneous emission. At the temperature of most planets, $N_l$ greatly exceeds $N_u$ in LTE, so the negative term dominates, and absorption is observed. If the level populations were not in LTE, or if the planet were very hot, then the spontaneous emission term ($A_{lu} N_u$) could become important \citep{kipping10}, and it would reduce the magnitude of the absorption to the point where it would have to be accounted for in analyses of transit spectra.  The most interesting case (not yet observed during transit) is when $N_u$ exceeds $N_l$, i.e., there is an inversion in level populations.  That requires unusual circumstances, but it does occur for carbon dioxide in the mesosphere of Mars \citep{deming_mumma}, and could conceivably also happen in some exoplanets.  When there is a level inversion, then stimulated emission (the $B_{ul}{N_u}$ term) dominates, and the exoplanetary atmosphere would {\it amplify} stellar radiation, rather than absorb it. Not only would that produce a spectacular transit emission spectrum, but it would also give us considerable insight into the atomic/molecular physics of the exoplanetary atmosphere. The circumstances that could produce a level inversion are beyond the scope of this tutorial, but if the phenomenon occurs in exoplanetary atmospheres it will probably be revealed at thermal infrared wavelengths.

Transit spectra refer to the terminator of the planet, where the line of sight passes through both day and night hemispheres.  Consequently there can be considerable variation in temperature, density, and cloud conditions along the line of sight.  Nevertheless, a single atmospheric temperature profile is often used to model transit spectra, even isothermal atmospheres are sometimes used.  There are two reasons for such crude approximations.  First, we don't know the real atmospheric conditions (although general circulation models could give us theoretical atmospheres), so we have little choice.  Second, the number densities and absorption are strongly peaked at the tangent altitude, so a single 1-D atmosphere is not such a bad approximation.  \citet{fortney10} studied the validity of a 1-D approximation using 3-D models, and they found that a 1-D approximation is generally valid, but it can be poor if the temperature is near transition points between different molecular compositions.  \citet{kempton14} cautioned that atmospheric winds can limit the applicabilty of 1-D models when high spectral resolution is used, and \citet{line_parmentier} studied degeneracies due to patchy clouds, and advocated modeling transmission spectra using 3-D general circulation models. 

Clouds in the exoplanetary atmosphere are a major effect that must be considered in the interpretation of transit spectra.  This was apparent from the first detection of an exoplanetary atmosphere by \citet{charbonneau02}.  Those authors noted the weakness of atomic sodium during transit, and suggested high clouds as the mostly likely explanation. Similarly, the first WFC3 spatial scan spectra \citep{deming13} showed water vapor absorption that was substantially weaker than would be produced by a clear atmosphere.  \citet{deming13} noted that adding clouds to the modeled atmosphere is much more effective at reducing the observed strength of water absorption than is reducing the water abundance.  There is a degeneracy between cloud height and abundance (not just for water vapor or sodium, but generally).  Clouds that block the lower atmosphere, combined with a high abundance of absorbers above the clouds, can produce approximately the same transit spectrum as a low abundance of absorbers in a clear atmosphere.  That degeneracy is illustrated for a hypothetical spectrum of the mini-Neptune GJ1214b in Figure~\ref{degeneracy}. Fortunately, the degeneracy can be broken by using the detailed shape of spectral features (e.g., the presence of pressure-broadened wings of absorption), and by including data at other wavelengths.  The most complete analyses use a retrieval approach on the total observations to break degeneracies (see below).

\begin{figure}
\includegraphics[width=3in]{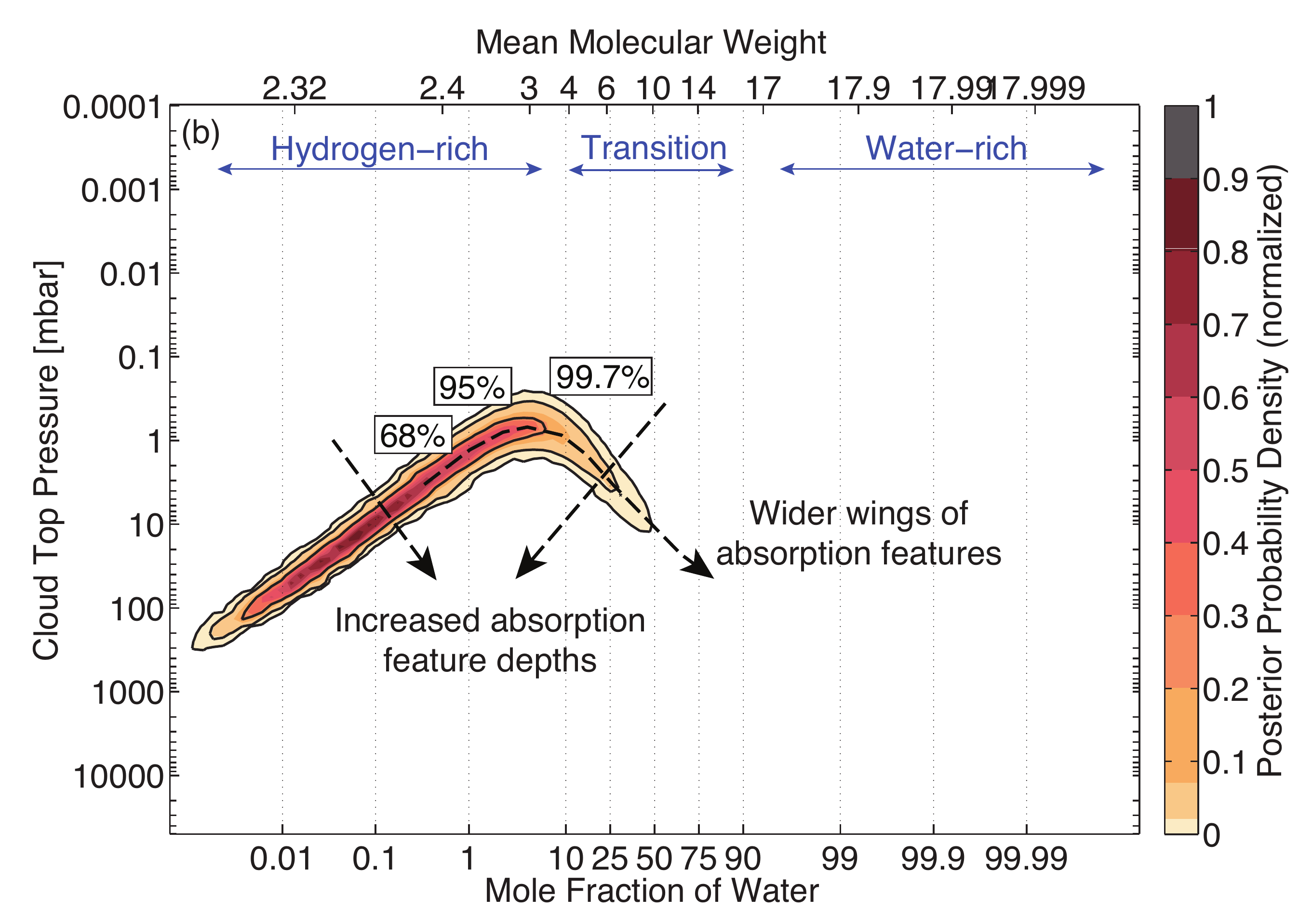}
\caption{Correlated posterior distributions for water vapor mixing ratio (X-axis) and cloud top pressure (Y-axis) for hypothetical transit spectra of the mini-Neptune GJ1214b, from \citet{benneke13}. Over a range of low abundance, the retrieved water mixing ratio correlated with the cloud top pressure; the transit spectrum can be matched by a small mixing ratio and cloud tops at high pressure, or a higher mixing ratio with cloud tops at greater pressure (red shaded region sloping up and the the right). }
\label{degeneracy}
\end{figure}

The cloud-abundance degeneracy is a special case of a more general degeneracy involving the 'reference pressure' \citep{griffith14, heng17}.  The concept is that the magnitude of transit absorption can be expressed in terms of the number of atmospheric scale heights that are opaque in a specific spectral feature (Eq.~8), and similar observed spectra are produced whether those 'N scale heights' occur at low pressure or high pressure.  Unless we know the continuous opacity of the exoplanetary atmosphere (including clouds) very accurately, any given amount of absorption during transit could potentially be ascribed to a low relative abundance of absorbers making the atmosphere opaque over N scale heights at high pressure, or a high relative abundance of absorbers combined with a high continuous opacity, shifting the N opaque scale heights to lower pressure, but producing closely the same transit absorption spectrum.  In principle, the pressure at the absorbing layer could be inferred from the profiles of individual atomic or molecular lines that are pressure-broadened.  In practice, that is usually only possible using the strong fundamental lines of sodium and potassium.  \citet{nikolov18} used the wings of those lines in the hot-Saturn WASP-96b to demonstrate that the atmosphere of that planet was free of clouds, and thus break the degeneracy and determine the abundance of free sodium. 

Some minor effects (compared to clouds) that should be considered in the interpretation of transit spectra include refraction \citep{hui02, betremieux15, misra14a}  and scattering \citep{robinson17}.  Refraction bends our line of sight through the exoplanetary atmosphere, and it can re-direct stellar light to pass through the exoplanetary atmosphere even outside of transit, producing 'shoulders' on the transit curve \citep{dalba17}.  Refraction is of potentially high interest because refraction effects are most prominent in clear atmospheres with a large scale height \citep{sidis10, misra14b}, exactly the circumstances that produce strong transit absorption signals, thus marking potential prime targets for transit spectroscopy.  Unfortunately, the presence of clouds weakens the amplitude of the refraction shoulders, believed to be no greater than a few parts per million (ppm, \citealp{alp18}), making them very difficult to detect.  An important but indirect effect of refraction is that it produces an upper limit to the pressures that can be probed by transit spectroscopy, even in clear atmospheres \citep{garcia12}. Similarly, scattering can increase the effective opacity of the atmosphere, thereby affecting the strength of spectral absorption during transit by as much as 200 ppm \citep{robinson17}. 

\subsubsection{Interpreting eclipse Spectra}

Spectra of transiting exoplanets can be measured at secondary eclipse \citep{richardson07, grillmair08, swain08, todorov14}.  However, eclipse spectra are (so far) relatively few since only the hottest planets emit at HST bands \citep{haynes15, sheppard17, arcangeli18, kreidberg18}, and Spitzer provides photometry, but does not resolve molecular features such as carbon monoxide. The scarcity of eclipse spectra will change when JWST begins to observe secondary eclipses \citep{greene16}. The interpretation of eclipse spectra differs fundamentally from transit spectra.  Whereas transit spectra are not sensitive to the radiation emitted by the exoplanetary atmosphere, only to absorption of stellar radiation, eclipse spectra represent only exoplanetary radiation.  The emergent spectral intensity at wavelength $\lambda$ (and in the vertical direction) is given as \citep{rybicki, seager_book}:

\begin{equation}
I_{\lambda} = \int_{0}^{\infty} S(\tau) e^{-\tau} d\tau,
\end{equation}

where the integral is over optical depth, and $S(\tau)$ is the source function at optical depth $\tau$, and the source function equals the Planck function in LTE.  The relative abundance of a spectral absorber affects the optical depth at the wavelength where that species absorbs. The more of a given absorber, the more that optical depths near unity are shifted to lower column densities, higher in the atmosphere.  Optical depth (via $e^{-\tau} d\tau$) essentially weights the source function in the above expression for the emergent intensity.  Consequently, increasing the abundance of an absorbing species makes a greater absorption (less emergent intensity) in the (normal) case where the source function (i.e., Planck function) decreases at greater heights.  If the Planck function (via temperature) increases with increasing height, then an emission spectrum is produced.  The strength of absorption (or emission) features seen in eclipse spectra thus depends on both the relative abundance of the absorbing species, and also on the temperature in the atmosphere. Note, for example, that an isothermal atmosphere will have no spectral features when in LTE (source function = Planck function).  A pedagogical cookbook for radiative transfer calculations in exoplanetary atmospheres is given by \citet{heng_marley}. 

In principle, transit and eclipse spectra should be analyzed together, since they both refer to related aspects of the same planet. That would require relating the temperature profile on the entire day side atmosphere of the planet (affecting eclipse spectra) with the temperature at the terminator (affecting transit spectra). Note that the temperature at the terminator is a principal determinant of the atmospheric scale height, so transit spectra are sensitive to temperature even without being sensitive to the emergent radiation from the atmosphere.  Relating the temperature on the day side to temperature at the terminator could in principle be done using general circulation models, but we are unaware of any such attempts to date.  Moreover, the results would be sensitive to the uncertainties inherent in the circulation model.     

The first studies of exoplanetary spectra using transits and eclipses (e.g., \citealp{charbonneau02, knutson08, machalek09}) merely compared a few models of the eclipse depth, calculated as described above, directly to the observations.  But sophisticated 'retrieval' methods are now used \citep{madhu09, line13, line14, waldmann15, lavie17, oreshenko17}; retrievals are based on calculating a large number of exoplanetary models using combinations of parameters such as the temperature profile and the atmospheric composition \citep{madhu09}.  The retrieval process varies combinations of those parameters extensively, and defines the range of model combinations that are consistent with the observations.  The process of defining that range of models is usually based on Bayesian statistics, and retrievals implement efficient methods to search parameter space for plausible models. Those search methods include Markov Chain Monte Carlo \citep{goodman10} or nested sampling algorithms \citep{skilling04,lavie17}.  \citet{waldmann15} explored several Bayesian approaches to retrievals using machine learning.  \citet{lavie17} used nested sampling and Bayesian evidence to evaluate equilibrium versus disequilibrium chemistry.  Work by \citet{oreshenko17} demonstrated the importance of priors when interpreting sparse data sets via retrievals.  They showed that a debate concerning the carbon-to-oxygen ratio for the hot exoplanet WASP-12b reduces to a debate about assumptions on the priors of the volume mixing ratios or elemental abundances.  

An example of a retrieval of the HST + Spitzer secondary eclipse spectrum of the iconic hot Jupiter HD\,209458b by \citet{line16} is shown in Figure~\ref{retrieval_fig}.  That example illustrates that there are usually degeneracies in the interpretation of observations of exoplanetary atmospheres.  Many of those degeneracies will be broken using spectra of the quality anticipated from JWST \citep{batalha17, chapman17, oreshenko17, schlawin18}.   

\begin{figure*}[ht]
\includegraphics[width=7in]{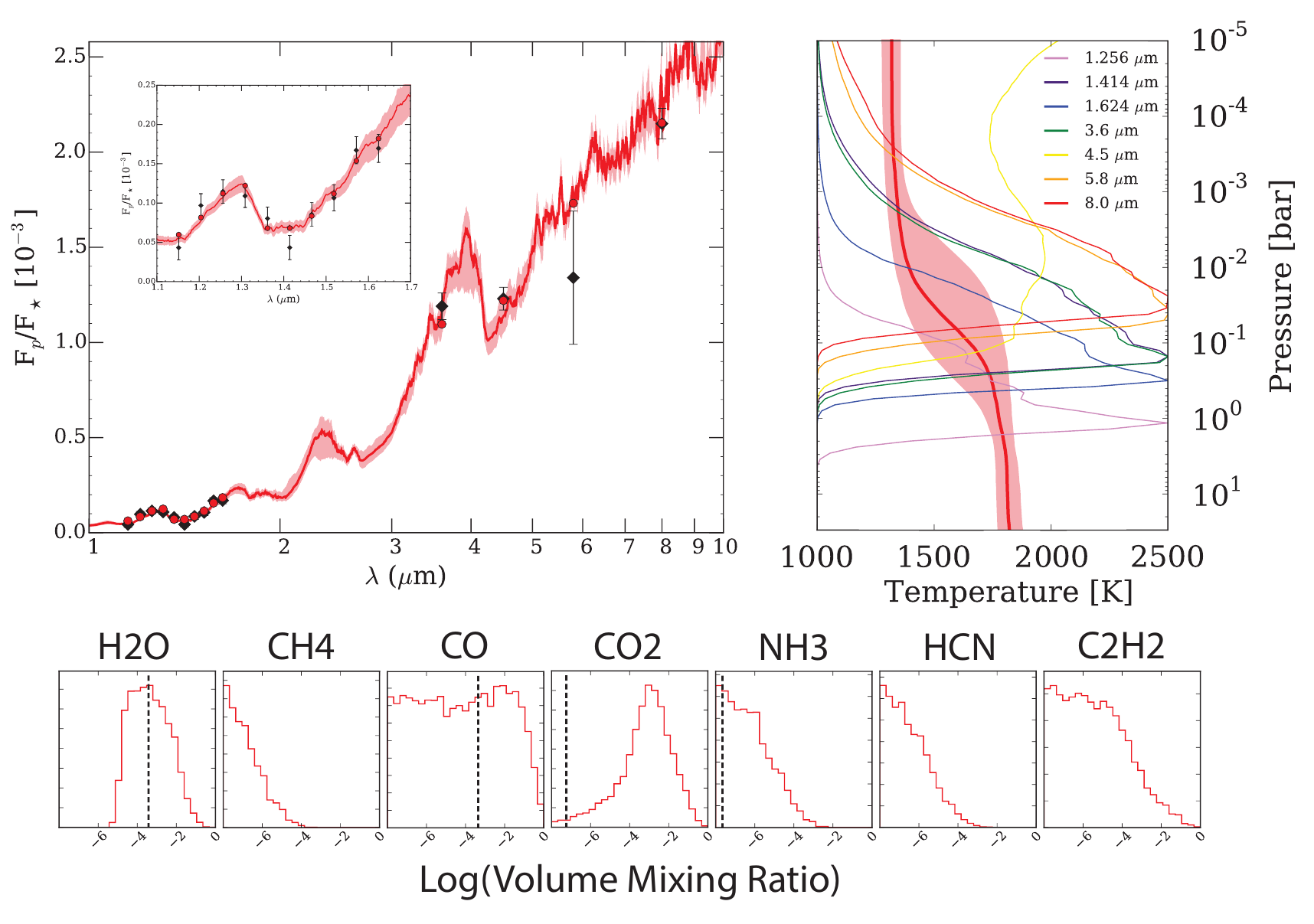}
\caption{Example of a retrieval analysis applied to the secondary eclipse spectrum of the hot Jupiter HD\,209458b by \citet{line16}. The upper left panel shows the emergent spectrum as measured by Spitzer photometry, and HST/WFC3 spectroscopy (inset).  The upper right panel shows the span of allowed temperature profiles (shaded red region is $\pm1-\sigma$ range) that are consistent with the measured data.  The other curves are contribution functions that express the relative contibution of each pressure level to the emergent radiation in the labeled bands.  The lower panels are the possible mixing ratios of various molecular absorbers.  Note that some mixing ratios (e.g., water vapor) are relatively well-constrained, while others (e.g., CO) could have a large range of values.  This illustrates the strength of retrieval analyses: they allow us to specify what parameters are consistent with the data, to what level of uncertainty. }
\label{retrieval_fig}
\end{figure*}

It is important to note that retrievals are not {\it inversions}, in the sense that the observational data are not manipulated to directly calculate the state of the exoplanetary atmosphere.  A direct inversion of that type is an ill-posed problem for exoplanetary atmospheres, because there would be an infinite number of solutions \citep{backus67}. Retrievals compare a large collection of forward models to the data in order to define the range of parameters that characterize that infinite space of solutions.

Finally, we point out one aspect that has received insufficient attention when analyzing eclipse spectra.  That neglected aspect is the effect of temperature inhomogeneities over the disk of the planets (i.e., as a function of latitude/longitude). Inhomgeneities can be caused by small-scale clearings in clouds, that permit radiation to well up from the deeper atmosphere, such as seen on Jupiter at 4- to 5\,$\mu$m \citep{grassi17}.  At wavelengths shortward of the peak of the Planck function, the emitted flux depends on temperature in a strongly non-linear fashion.  Hence the emitted flux averaged over the disk of the planet may not accurately reflect the average temperature, and temperature profiles derived by retrievals could in principle be biased by this effect.  \citet{wilkins14} raised the issue of possible temperature inhomgeneities when interpreting the secondary eclipse spectrum of CoRoT-2b, and \citet{feng16} did some pioneering modeling of temperature inhomogeneities, but there has been relatively little additional work on this important topic.

\subsection{Statistical Studies} 
\label{subsec:statistical}

The success of the ground-based transit surveys, and Kepler and (soon) TESS \citep{ricker15}, have provided us with a very large number of known transiting exoplanets.  We already have more transiting exoplanets than can be studied in practice using the premier facilities that are required for atmospheric characterization (e.g., Hubble, JWST, ELTs).  Moreover, many of these planets are sufficiently distant from Earth that the signal-to-noise for characterizing their atmosphere would be marginal.  The general solution to this problem of too many planets and not enough signal-to-noise, is to approach atmospheric characterization statistically.  The value of statistical studies has been argued by \citet{bean17}, who point out the value of 'performing surveys of key planetary characteristics and using statistical marginalization to answer broader questions than can be addressed with a small sample of objects.'  A simple example of a statistical study that yields information about hot Jupiters is the work of \citet{triaud14a, triaud14b}, who studied the colors of hot Jupiters using their fluxes measured from secondary eclipses (also, see \citealp{beatty14}).  Secondary eclipses are well suited to statistical studies, because they can be observed in only a few hours per eclipse, and both Spitzer \citep{garhart18b} and ground-based telesopes \citep{martioli18} have good sensitivity for eclipses of hot Jupiters.

\begin{figure}
\includegraphics[width=3in]{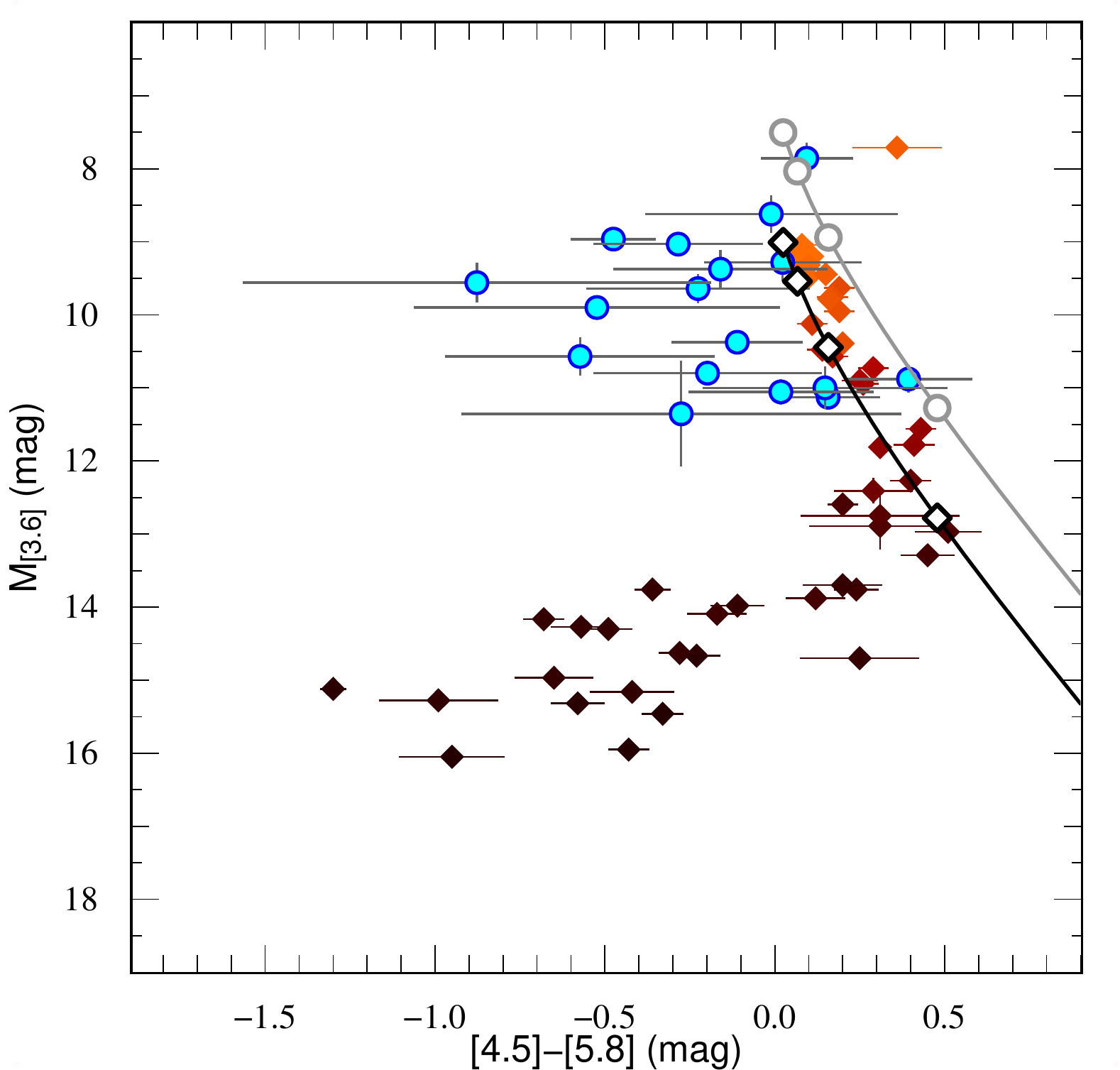}
\caption{Statistical analysis of hot Jupiters by studying their color (relative brightness in Spitzer bandpasses, as measured using secondary eclipses).  The X-axis is the magnitude difference in the 4.5 and 5.8\,$\mu$m Spitzer bands versus the absolute magnitude of the planet in Spitzer's 3.6\,$\mu$m band on the Y-axis.  Hot Jupiters are the light blue symbols; the orange to black symbols are the sequence of low mass stars, from M5 to Y1 spectral types.  The lines are the locii of blackbodies having 0.9 (black line) and 1.8 (gray line) Jupiter radii.  The diamonds along the blackbody lines mark temperatures of 4000, 3000, 2000, and 1000K.  This Figure reproduced from \citet{triaud14b}.}
\label{triaud_fig}
\end{figure}

Figure~\ref{triaud_fig} from \citet{triaud14b} shows the color of transiting hot Jupiters measured by comparing two Spitzer bands (4.5 and 5.8\,$\mu$m), plotted versus the absolute magnitude of the planet at 3.6\,$\mu$m.  Compared to low mass stars and brown dwarfs, hot Jupiters (blue triangles) show a wider range of color at the same absolute magnitude, and they scatter to left of the locus defined by low-mass stars and blackbodies.  That behavior can be produced by extra absorptions in the Spitzer bands, produced by molecules in the exoplanetary atmosphere.  Scatter to the left indicates either extra absorption at 5.8\,$\mu$m, or a lack of absorption (or even, emission) in the 4.5\,$\mu$m band. 

Another example of how secondary eclipses can be used statistically, is the study of longitudinal heat re-distribution. \citet{cowan11} pointed out that significant constraints on heat re-distribution can be obtained by studying the day side brightness temperaures of a collection of hot giant planets, and relating those temperatures as measured using secondary eclipses versus the maximum day side temperature.  For example, if measured day side temperatures are consistently less than the maximum possible temperature by a specific factor, then either heat is being re-distributed longitudinally, or the planets have reflected stellar irradiance to a greater degree (higher albedo) than is commonlly believed.  An example of this relation is shown on Figure~\ref{longitudinal}, from the large secondary eclipse study by \citet{garhart18b}. Horizontal lines on this relation give the cases of complete longitudinal distribution of heat (dotted line), redistribution over the day side hemisphere (dashed line), and no redistribution (solid line).  Planets can fall below the dotted line if their albedo exceeds zero.  But the hottest planets should have low albedos (clouds cannot condense), and indeed above $\sim 2300$K most planets lie primarily between the regimes of re-distribution over the day side hemisphere and complete longitudinal redistribution (dotted line).  The fact that none fall below the dotted line is statistically consistent with very low albedos for the hottest giant planets.

\begin{figure}
\includegraphics[width=3in]{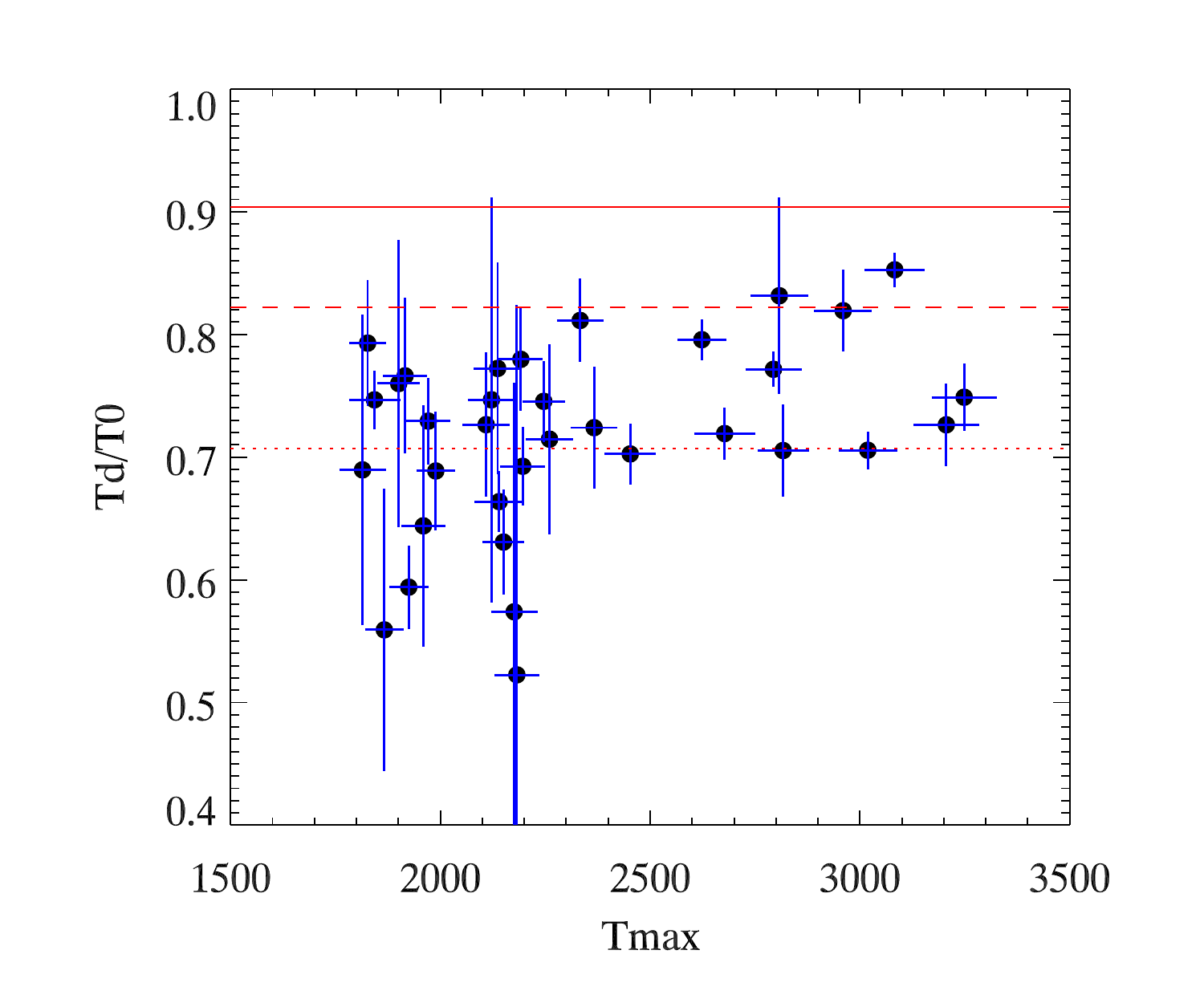}
\caption{Day side temperature (Y-axis), measured from secondary eclipses and normalized by the equilibrium temperature at the sub-stellar point for zero albedo ($T_0$), plotted versus the maximum possible temperature for the day side hemisphere (X-axis).  The dotted, dashed, and solid red horizontal lines are the temperatures for complete heat redistribution, day side hemisphere redistribution, and no redistribution, respectively.  This Figure is from \citet{garhart18b}, updated with new data from an original version by \citet{cowan11}. }
\label{longitudinal}
\end{figure}

\subsection{Stacking} 
\label{subsec:stacking}

Another statistical technique that is just beginning to be used for exoplanetary characterization is stacking.  Stacking means that signals such as secondary eclipses for {\it different planets} are scaled and appropriately shifted and co-added.  This allows the detection of weak events for a group of planets whose signal-to-noise would not be adequate to detect individually.  Stacking has been widely used in astrophysics (e.g., \citealp{niikura15}), but only a few papers have exploited the stacking of exoplanet data.  \citet{sheets14} and \citet{sheets17} used stacking to measure the reflected light from Kepler planets as a function of radius, by stacking tens of thousands of secondary eclipses, and derived geometric albedos as shown in Figure~\ref{stack_fig}. They find that small planets are dark, with average geometric albedos near 0.1 for planets with radii between one and four Earth radii, as shown in Figure~\ref{stack_fig}.  Similar low albedos were found for an ensemble of Kepler planets by \citet{jansen18}, who stacked phase curves for planets cooler than the planets analyzed by \citet{sheets17}.  Beyond atmospheres, \citet{hippke15} stacked Kepler transit data to search for a population of small bodies (comets, asteroids) at stable Trojan points in orbits of Kepler planets.

\begin{figure}[ht]
\includegraphics[width=3in]{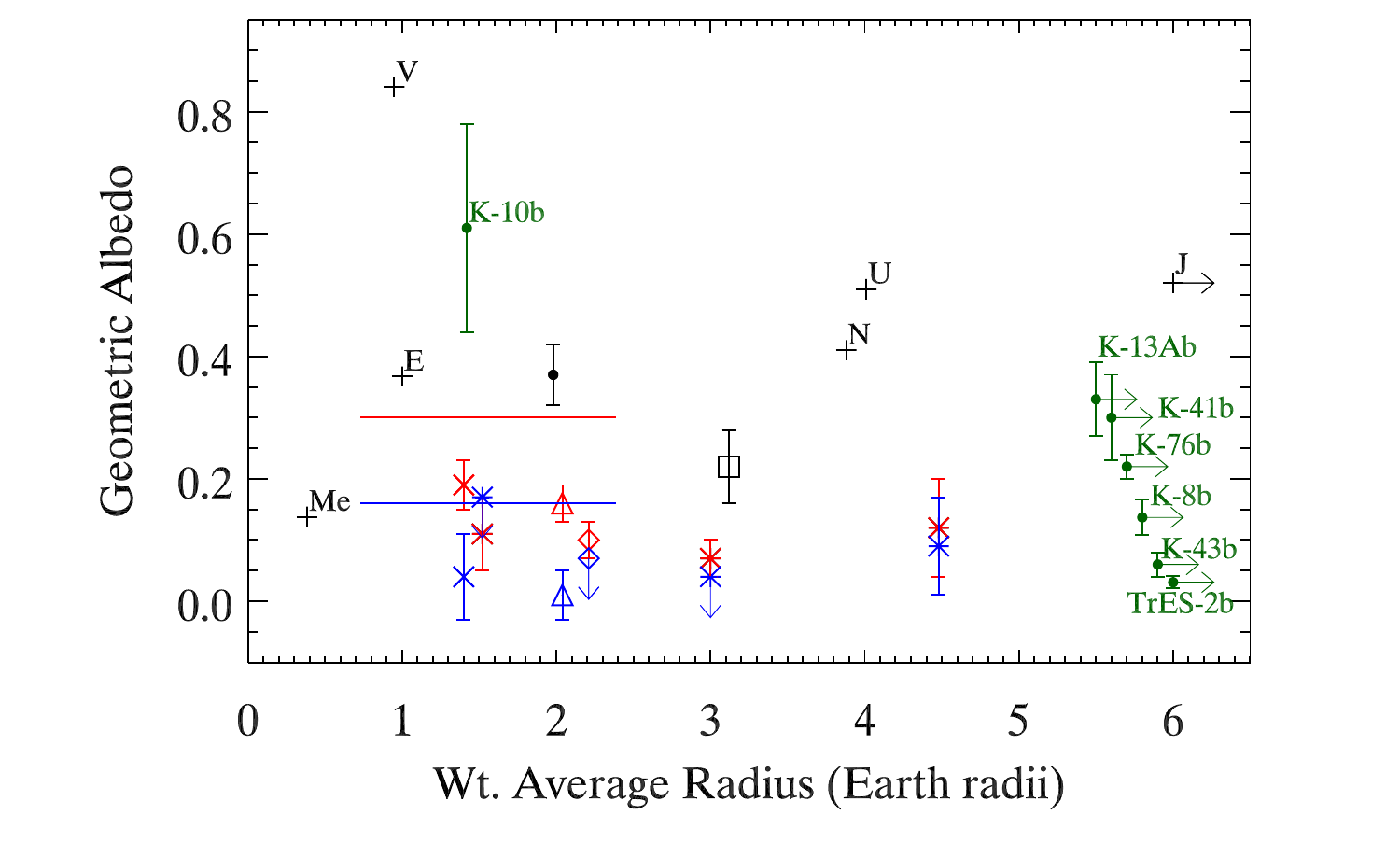}
\caption{Statistical results for geometric albedos of Kepler planets via the technique of stacking secondary eclipses of multiple planets, from \citet{sheets17}. The red and blue points are the results for super-Earths, mini-Neptunes, and super-Neptunes, with different degrees of adopted longitudinal heat re-distribution (when subtracting thermal emission to infer reflected light).  The median albedos found in a Kepler eclipse study by \citet{demory14} are shown as horizonal red and blue lines.  Some planets with signal-to-noise that allows individual analyses are also illustrated, such as Kepler-10, and planets in our Solar System are given as letters.  Albedos from the literature for hot Jupiters are also illustrated (radii off scale to the right).}
\label{stack_fig}
\end{figure}

Stacking is a powerful technique because it yields information that would otherwise be unobtainable.  But that information is potentially ambiguous, because it refers to a group of planets, not individual worlds.  If the distribution of underlying properties is (for example) bimodal, then the result for the group may not be representative of any individual world.  Moreover, stacking is sensitive to several subtle sources of bias, and must be used with great care.  Investigators who wish to reap the considerable benefits of the stacking method are advised to study the analysis described in \citet{sheets17}.
 
\section{Effect of Stellar Activity} \label{sec:activity}

Magnetic activity on the host star can interfere with our ability to derive robust exoplanetary spectra, especially transmission spectra during transit.  Although we derive the transmission spectrum as a relative measurement in which the stellar flux appears to cancel (see Sec.~2.4), the planet itself ruins the purely relative nature of the measurement.  When the disk of the star is featureless, and the stellar flux is constant with time, then the equations given in Sec.~\ref{subsec:transits} apply strictly. In that case, deriving a robust exoplanetary spectrum is just a matter of having sufficient signal-to-noise, and controlling systematic errors from the instrumentation.  However, real solar-type (FGK) or especially M-dwarf stars seldom have featureless disks.  When the stellar disk is spatially inhomogenous, then the blocking by the planet can cause the stellar flux during transit to have a different spectral character than outside of transit. For example, star spots often occur at 'active latitudes' displaced from the stellar equator.  If the planet orbits in the star's equatorial plane, then it will transit on the stellar equator, and may never block any star spots. Hence the stellar spectrum during transit will have a greater relative contribution from the spots, that are cool and exhibit molecular absorption features (e.g., water vapor, \citealp{wallace95}).  Moreover, that increase in the prominence of molecular features will be temporally synchronous with the transit, and could thus be interpreted as being due to the absorption spectrum of the planet.  The effect of unocculted star spots has been known for at least a decade \citep{pont08}.  \citet{deming13} demonstrated that it was negligible for the case of the solar-type - and magnetically quiet - star HD\,209458b, and they give some formulae that can be used to estimate the magnitude of the effect.  \citet{mccullough14} modeled the effect of unocculted star spots for the active star HD\,189733b, and they find that unocculted spots can mimic the effect of a scattering haze \citep{pont13} in the exoplanetary atmosphere.  Both effects make the planet's transit radius appear to increase at short wavelength. The scattering haze does that by increasing the opacity of the atmosphere at shorter wavelength, whereas unocculted star spots make the planet look bigger by reducing the stellar flux at short wavelength.  A comparison of the two effects is shown on Figure~\ref{189_fig}, from \citet{mccullough14}.
 
\begin{figure*}
\includegraphics[width=5in]{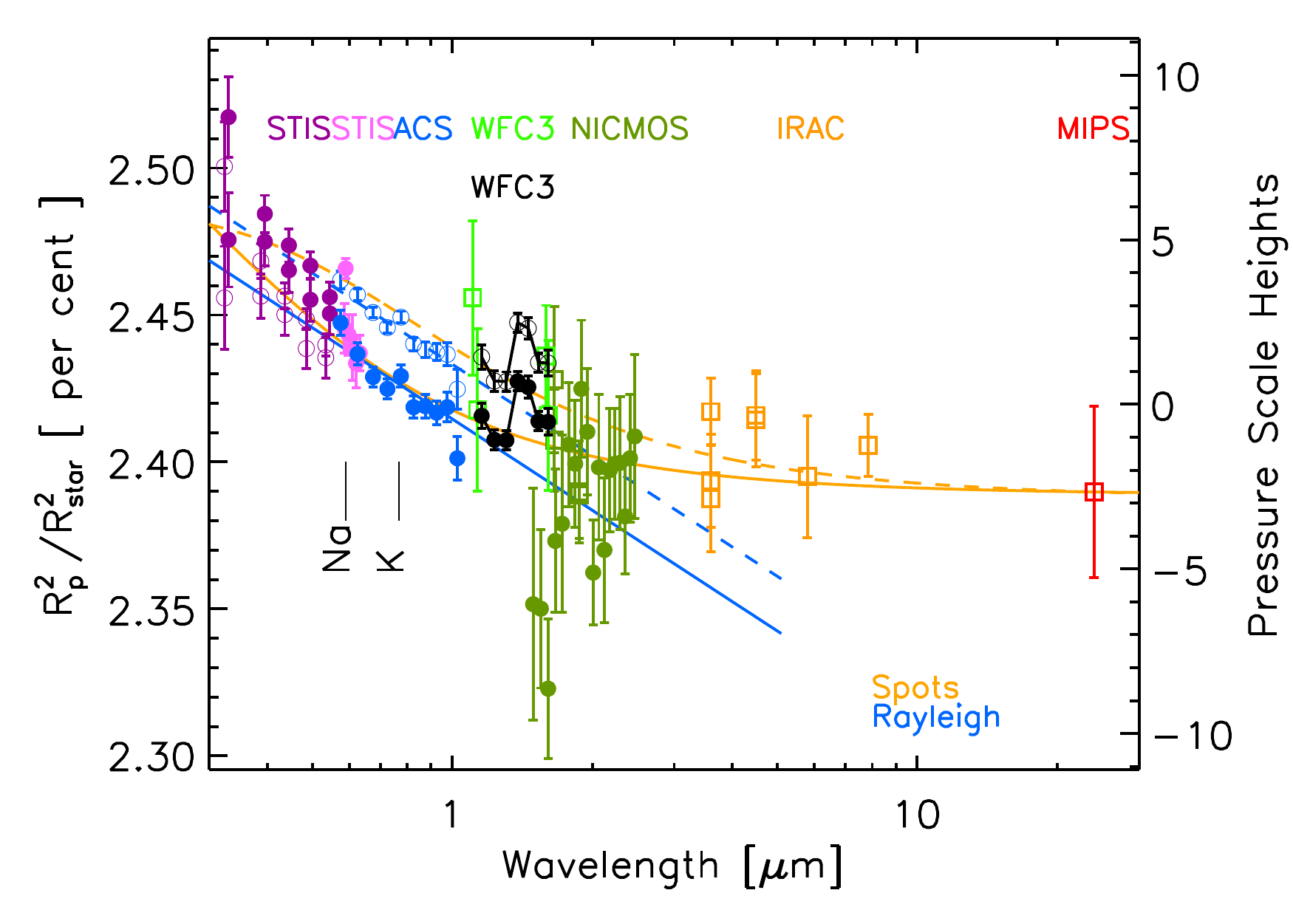}
\caption{Effect of unocculted star spots on the transit spectrum of the hot Jupiter HD\,189733b, from \citet{mccullough14}.  Star spots are cooler than the stellar photosphere; hence a transiting planet that preferentially blocks non-spot regions of the photosphere will make the star look cooler during transit.  The stellar flux during transit thereby decreases with decreasing wavelength, making the planet appear larger at shorter wavelengths.  The net result is very similar to the effect of Rayleigh scattering by small particles in the exoplanetary atmosphere.  Other spectral features (e.g., atomic Na and K resonance lines) can be used to help determine which interpretation is correct.}
\label{189_fig}
\end{figure*}

Planets can also cross spots during transit, and sufficiently large spots will cause noticeable perturbations to the transit curve \citep{rabus09}.  Spot crossings cause the flux to {\it increase}, because the spots are dark, so the planet is not blocking as many photons.  The crossing of large spots can be readily identified in transit photometry \citep{morris17}, but small spot-crossings may go unnoticed.  The cumulative effect of small spot crossings will be to make the planet appear smaller, especially at shorter wavelengths.  In principle, that could mask real absorption effects in the exoplanetary atmosphere.  Moreover, stellar activity during transit is not confined to star spots; stellar disks also exhibit bright plage areas, as well as convective patterns such as granulation and super-granulation.  The total effect on exoplanet transit curves can be quite complex, especially for M-dwarf stars \citep{rackham18}.  In addition to the effects from geometric blocking of the stellar disk, there is also a spectral resolution-dependent bias in transit spectroscopy that becomes significant when observing strong spectral features common to the star and planet (e.g., water vapor for M-dwarf planets). Observations at low- to moderate spectral resolving power can allow flux-leakage between the stellar and planetary spectrum that will bias the inferred spectrum of the planet \citep{deming17}. 

Stellar activity is of less concern for secondary eclipses, because the planet passes behind the star and does not block portions of the stellar disk.  In that case (and also for transits), the possible temporal variability of the star must be considered.  Fortunately, stable stars (i.e., not large amplitude pulsators) tend to vary on significantly different time scales than planetary transits. Active stars usually have star spots.  As a spotted star rotates, the flux we measure using photometry varies when some regions of the star have more spots than other regions.  Fortunately, the rotation of a spotted star usually occurs on a longer time scale than a secondary eclipse or transit, and impulsive events such as stellar flares occur on shorter times scales than the transit.  For example \citet{seager09a} identified and successfully corrected a prominent stellar flare in their observations of GJ876, although a spectrum of lower amplitude flares could be less separable, and thereby a problem for transit measurements \citep{davenport17}.

The first step to correct transit observations for the effect of stellar activity is to verify that the star is active, and determine to what degree, either by photometric variability or spectroscopy of emission in the H \& K lines.   Photometric variability underestimates the degree to which spots occur on a star, because as one spot is rotating onto the Earth-facing hemisphere, others spots are rotating off, reducing the amplitude of variability in disk-integrated light. Estimating the nature and magnitude of stellar activity is a difficult problem, and a sub-field of astronomy in its own right.  \citet{fabian17} give a recent review of activity in cool stars, and some examples of innovative work to measure stellar activity are \citet{oneal06}, \citet{hall07}, \citet{crossfield14}, \citet{rottenbacher17}, and \citet{morris18}.    

\section{The Future} \label{sec:future}

As more powerful astronomical facilities come into operation, the principal benefits to exoplanet characterization will (arguably) come from spectroscopy.  While spectroscopy is easier to interpret than photometry (Sec.~\ref{sec:photspec}), it is harder to measure, requiring collection of many photons.  New large-aperture facilites such as the James Webb Space Telescope (JWST) and the ground-based Extremely Large Telescopes (ELTs) will collect the large photon fluxes that spectroscopy of exoplanetary atmospheres requires. 

\subsection{The James Webb Space Telescope} \label{subsec:JWST}

JWST is a spectroscopic powerhouse, with many spectroscopic modes that will characterize a wide range of transiting exoplanets, from hot Jupiters down to super-Earths. Numerous simulation studies have projected the results from JWST for exoplanets from hot Jupiters to habitable super-Earths \citep{seager09b, deming09, kaltenegger09, belu11, shabram11, beichman14, barstow15, greene16, kreidberg16, schwieterman16, arney17, morley17, batalha18}. Nevertheless, aspects of the JWST's future results will probably be surprising, because the potential for new phenomena in astronomy is vast.  We here discuss a major exoplanetary theme that JWST can address: the nature of the transition in atmospheres from Neptunes to rocky super-Earths, and the molecular composition of rocky super-earth atmospheres.

\begin{figure*}
\includegraphics[width=6in]{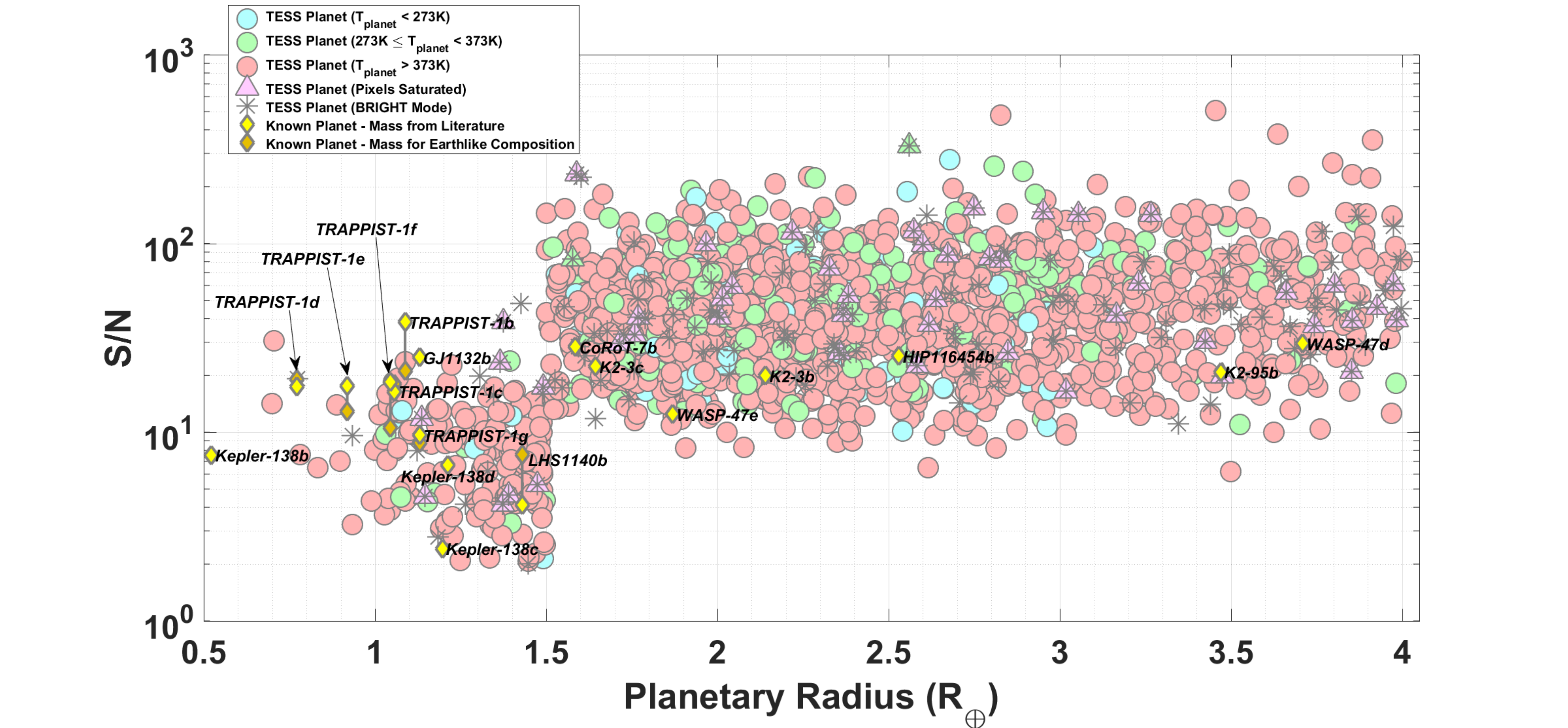}
\caption{Signal to noise for water vapor absorption of anticipated TESS planet discoveries, versus planetary radius. These calculations are from \citet{louie18}, and apply to observations using JWST for 10-hours per planet (only a relatively few planets will be observed).  Planets larger than $1.5 R_{\oplus}$ were modeled using solar composition (H-rich) atmospheres, whereas pure water vapor atmospheres were used for smaller planets.  That discontinuity in atmospheric composition causes the corresponding discontinuity in signal-to-noise shown on the Figure.  These signal-to-noise values assume clear atmospheres, and they apply to the integral of water vapor observed over the full NIRISS bandpass, not at any specific wavelength.
}
\label{louie_fig}
\end{figure*}

Masses and radii of transiting exoplanets yield their bulk density \citep{charbonneau00}, and we know that bulk densities increase as radius decreases, so that mini-Neptunes and super-Earths have higher bulk densities than do gas giants.  What we don't know is to what extent this applies to the {\it atmospheres} of exoplanets, i.e., to what extent the atmospheres of super-Earths have higher molecular weights than the hydrogen-dominated atmospheres of gas giants.  The abundance of elements heavier than helium is called the metallicity of the atmosphere, in analogy with the terminology used for stellar atmospheres.  One way to distinguish high metallicity exoplanetary atmospheres from hydrogen-dominated atmospheres, is by measuring the atmospheric scale height using transmission spectroscopy \citep{miller-ricci09}. The scale height is inversely proportional to the mean molecular weight in the atmosphere. During transit, the scale height is not measured directly, rather it is inferred from the strength of the absorption signal.  High metallicity (= high molecular weight) atmospheres will have small scale heights and small absorption signal during transit, whereas gas giants will have much greater scale heights and large absorption signals.  Since clouds will also decrease the signals, the scale height will in many cases have to be estimated using retrieval methods.  However, in some instances it may be possible to infer the scale height of the atmosphere robustly from a simple interpretation of the data.  For example, young super-Earths might host hydrogen-rich primordial atmospheres (molecular weight $\mu \sim 2.3$) that produce a transit absorption signal via 2 opaque scale heights (the absorber would be {\it trace} amounts of a heavier molecule).  In such cases, interpreting the result as being from a high molecular weight atmosphere (e.g., predominately carbon dioxide, $\mu =44$) could require more than an order of magnitude increase in the number of opaque scale heights (from 2 to 38), which would be physically unreasonable. Density decreases exponentially with the height measured in scale heights, and an altitude of 38 scale heights would have a mass density too low to produce significant absorption. In such cases it would be an easy inference to conclude for a low molecular weight atmosphere on a super-Earth, although retrieval methods would still be required to extract the quantitative details.

As a caveat to the discussion above, note that the number of scale heights that are opaque during transit is a function of spectral resolution.  Very few observations will fully resolve an exoplanetary spectral line, so the 2 scale height example cited above is normally realistic.  However, if enough photons can be collected to enable line-resolved spectra, then the resolved cores of lines can be opaque over tens of scale heights.  In such cases, analytic formulae can break down.  For example, the derivation of \citet{heng17} utilized an isobaric assumption, that would be inapplicable at very high spectral resolving power.

Population studies (e.g., \citealp{fulton17}) have found that there is a paucity of planets with radii between $1.5$ and $2.0 {R}_{\oplus}$, suggesting that there are two distinct populations among the sub-Saturn planets: gas-rich Neptunes with radii exceeding $2.0 {R}_{\oplus}$, and rocky planets with radii below $\sim 1.5 {R}_{\oplus}$ \citep{rogers15}.  The TESS mission \citep{ricker15} will discover many new transiting planets of all radii \citep{sullivan15, barclay18}, and JWST will be able to probe the composition of atmospheres on both sides of this gap \citep{louie18}.  Figure~\ref{louie_fig} projects the signal-to-noise ratio for water vapor absorption observed in a 10-hour-per-planet program using the NIRISS instrument on JWST \citep{doyon12}, as calculated by \citet{louie18}.  In the Figure, the transition from low- to high-molecular weight atmospheres is assumed to be discontinuous, but real planets may exhibit a power law in the atmospheric metallicity versus exoplanetary mass or radius.  The zero-th order way to infer the nature of the transition will be to notice how the magnitude of the water vapor absorption varies as a function of exoplanetary radius, as explained above.  Beyond that simple method, retrieval studies \citep{barstow15} will define the metallicity, element ratios, and cloud properties of planets down to the size of super-Earths \citep{greene16}.  Note that a simple transition from low- to high molecular weight will not necessarily occur as radius decreases from mini-Neptunes to super-Earths.  Atmospheres with very high metallicity are plausible even for mini-Neptunes, and the high metallicity can cause interesting consequences for atmospheric chemistry, such as abundant carbon dioxide in a Neptune-mass planet \citep{moses13}.

Super-Earths are of special interest for JWST; we currently have very little information on the composition of their atmospheres. Super-earths that are potentially habitable have been a topic of intense interest, and we look toward TESS to find the best transiting cases.  \citet{deming09} considered the exoplanetary yield of TESS, and projected that JWST would be able to measure the temperature (via secondary eclipse), and detect the major atmospheric molecules (via transit spectroscopy), for one to four TESS super-Earths of habitable temperature, although that would require a large investment of JWST observing time \citep{cowan15}.   More accurate calculations have become possible with further development of JWST's instrumentation, and \citet{greene16} studied JWST's ability to characterize exoplanetary atmospheres.  Figure~\ref{greene_fig} shows their results for the contrasting cases of a warm ($T \sim 700$K) Neptune and a cool ($T \sim 500$K) super-Earth, each with a variety of atmospheres.  Solar-abundance atmospheres are dominated by hydrogen (as the Sun is).  The scale height of the atmosphere is inversely proportional to molecular weight, so those atmospheres are extended to great height and have a large cross-section for absorption during transit, producing a readily detected transit signature. The higher molecular weight atmospheres (1000 times solar metallicity for the Neptune, and the pure water vapor atmosphere for the super-Earth) have much smaller scale heights, harder to detect.  The detections also become more difficult when clouds are present in the exoplanetary atmosphere: even the warm Neptune could be difficult for the high metallicity case if clouds are added.  Nevertheless, we are optimistic that JWST will characterize atmospheres of small planets down to super-Earths, and not only for the hottest super-Earths but also for at least one whose atmospheric temperature approaches the habitable range.  Note that we do not expect the detection of biosignatures - that will occur only if we are very lucky \citep{deming_seager17}.

\begin{figure}
\includegraphics[width=3 in]{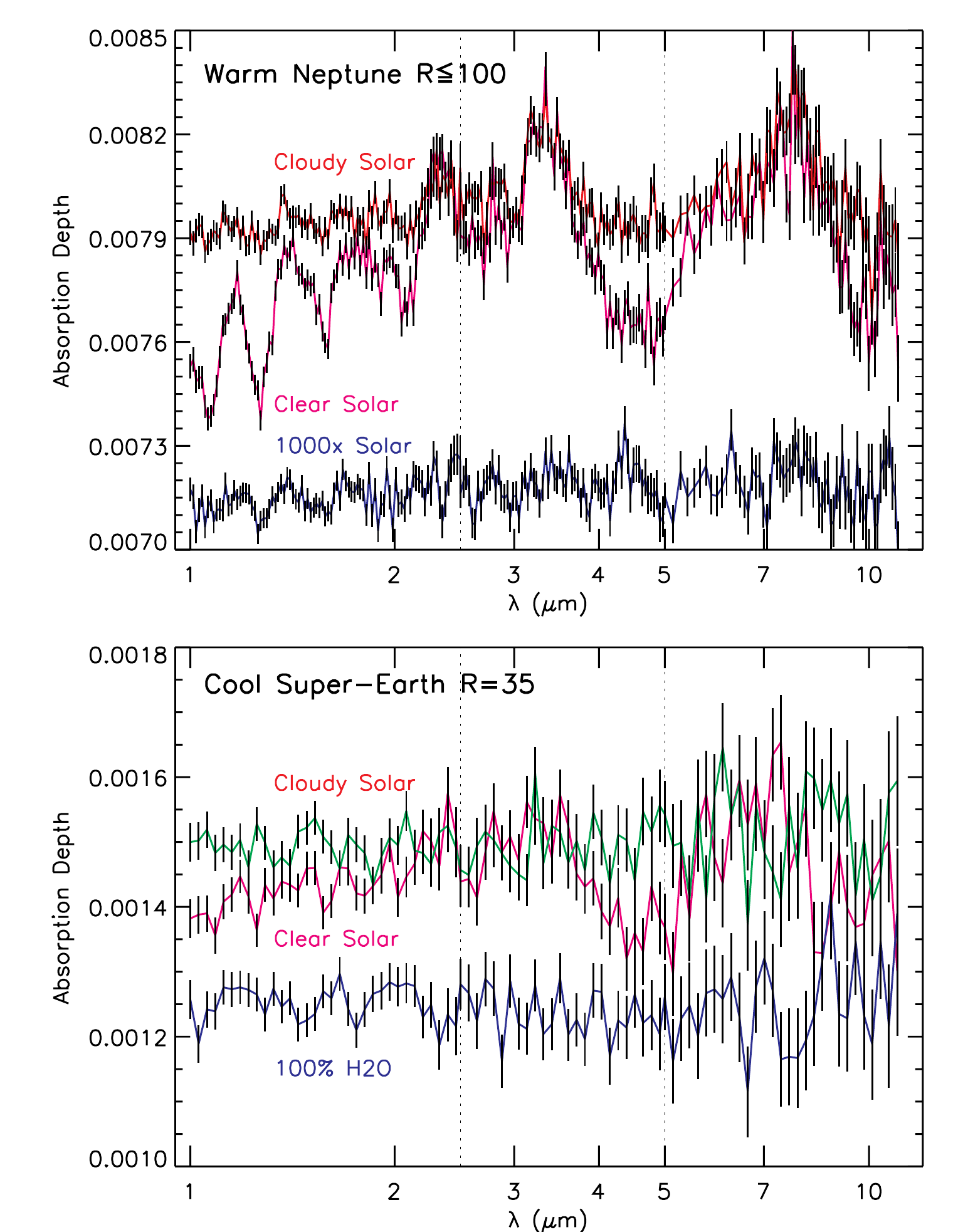}
\caption{Simulated JWST transmission spectra for a warm Neptune (top), and cool super-Earth (bottom), from \citet{greene16}. These spectra are for a single transit, and include simulated noise based on projected JWST performance.  Most of the modulation in the spectrum is due to water vapor absorption.
}
\label{greene_fig}
\end{figure}

\subsection{The Era of ELTs} 
\label{subsec:elts}

Although there is much excitement in the transiting exoplanet community concerning the prospects for JWST, the potential of very large ground-based telescopes is arguably greater.  One major advantage of JWST is that it operates cryogenically in the thermal infared, so it avoids the large thermal background flux (=noise) that strongly affect ground-based telescopes at long wavelengths.  However, transit spectroscopy is more favorable at short wavelengths due to the greater stellar flux, as we pointed out in Section~\ref{subsec:wavelength}.  At optical and near-infrared wavelengths, stars emit ample photons and a very large aperture ground-based telescope could collect an enormous photon flux.  If atmospheric noise can be cancelled or minimized, then the sensitivity could potentially be sufficient to detect atmospheric oxygen in a transiting Earth-like planet \citep{snellen13}. The large photon flux will make it practical to utilize high resolution spectroscopy.  The spectroscopic detection would utilize a cross-correlation technique (Section~\ref{subsec:convol}) wherein a modeled template spectrum is shifted in velocity and used as a numerical filter, multiplied times the target spectrum and integrated over wavelength.  Figure~\ref{snellen_fig} shows simulations of two transit spectroscopic detections of near-infrared molecular oxygen in the atmosphere of an Earth-like planet orbiting a small red dwarf star. There are several advantages of small stars for this type of detections \citep{charbonneau_deming07}.  Those advantages include a maximum signal-to-noise because there is minimal light from the star that doesn't contain the planet's absorption.  Moreover, orbital periods for planets orbiting low mass stars tend to be short (Kepler's law), so averaging 30 transits as in Figure~\ref{snellen_fig} can be accomplished in a reasonable time period.

\begin{figure*}
\plottwo{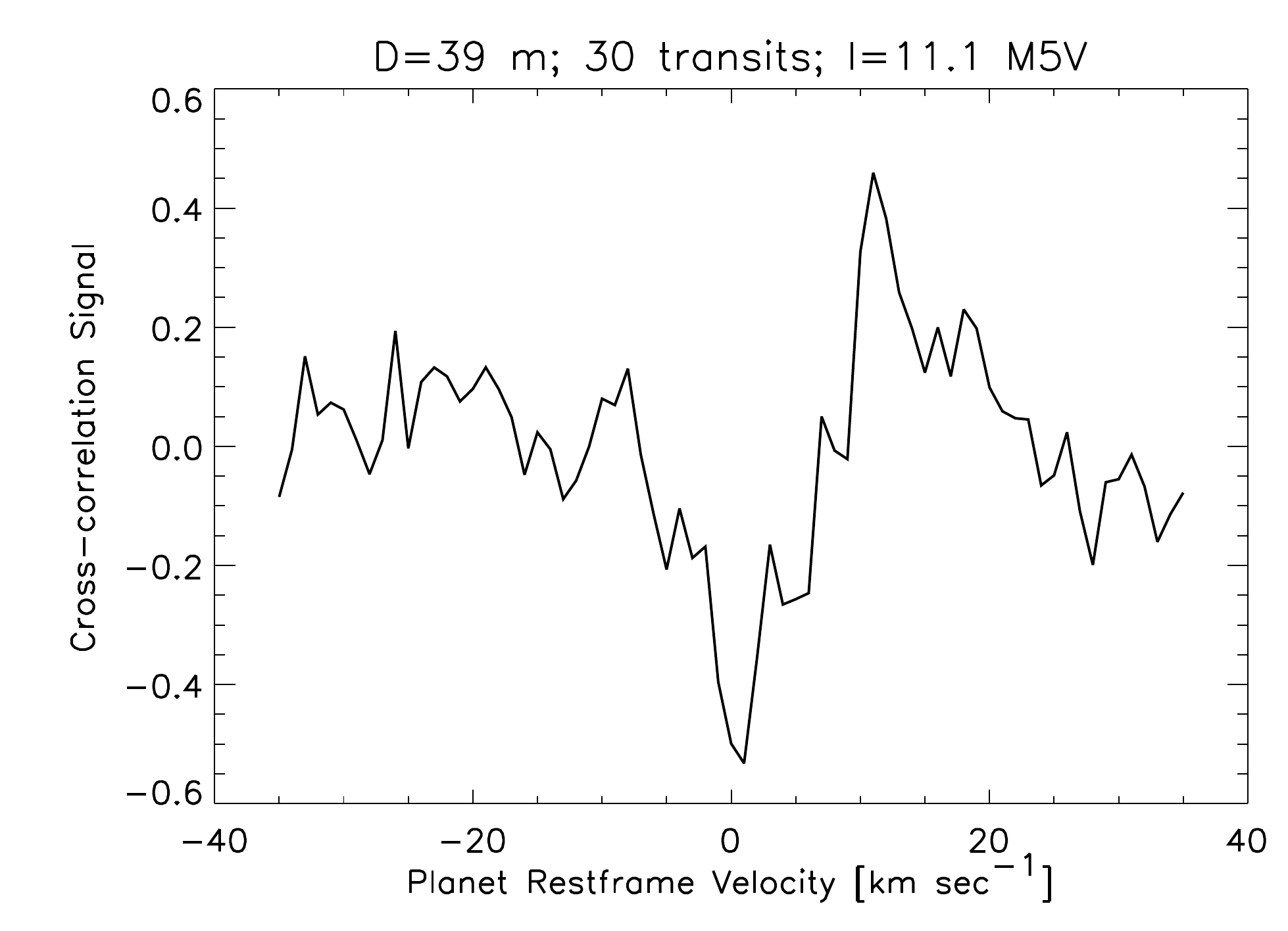}{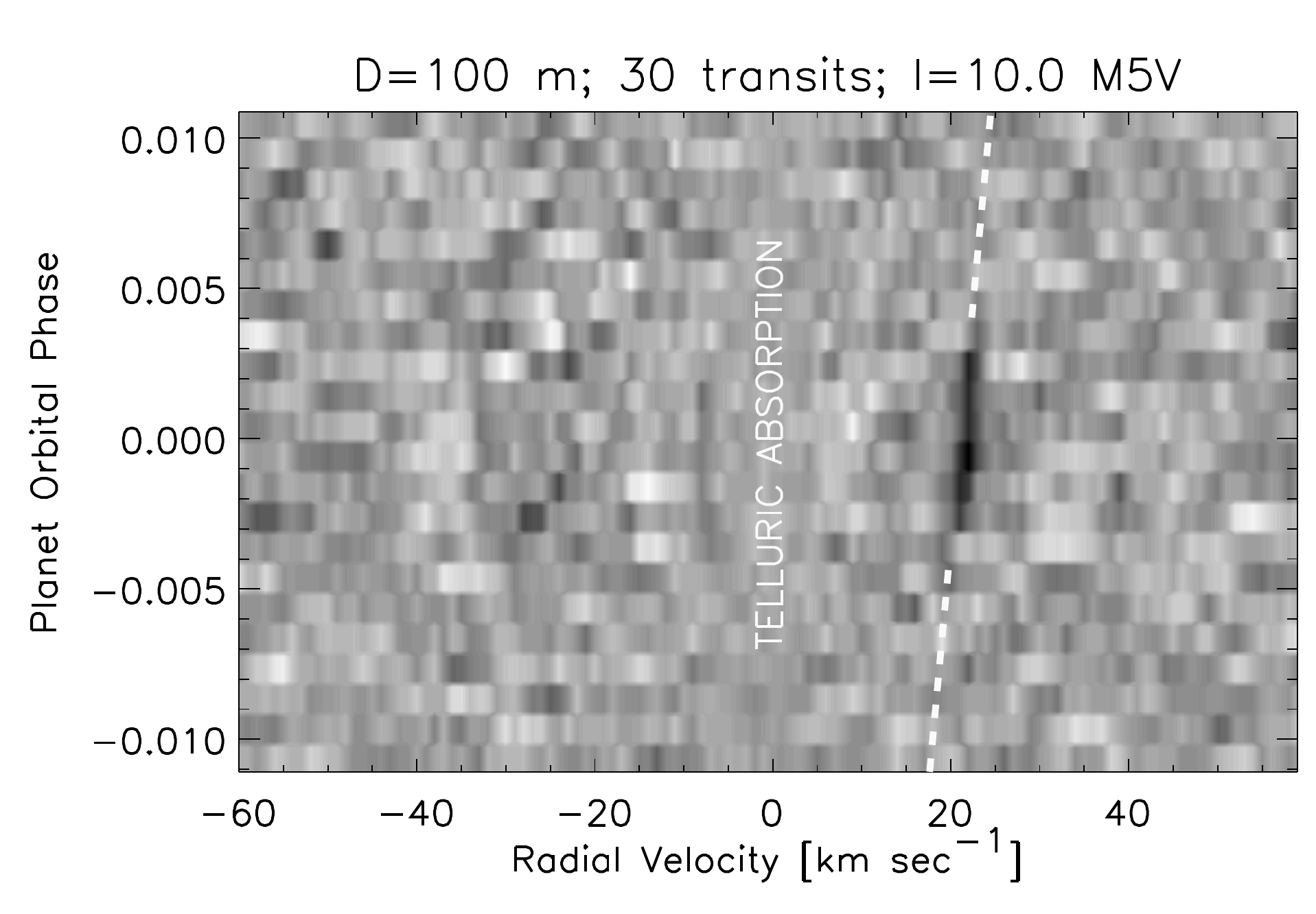}
\caption{Detection of molecular oxygen in the atmosphere of an Earth-like planet, at a wavelength of 1.27\,$\mu$m, using large-aperture ground-based telescopes, simulated by \citet{snellen13}.  The left panel shows a detection (dip at zero velocity in the rest frame of the planet), observed by a 39-meter telescope, and averaging over 30 transits.  The right panel shows higher signal-to-noise results also for 30 transits, but using a 100-meter telescope.  In the right panel, the Y-axis is the planet's orbital phase near transit, and the gray-scale indicates the strength of the signal.  The dark tilted line near 20 km/sec radial velocity is the oxygen absorption in the planet; it is tilted because the radial velocity of the planet is changing during transit.  (20~km/sec is the heliocentric radial velocity of the star+planet system).  In both cases, the host star is a cool red dwarf star, that is bright at this near-infrared wavelength.
 }
\label{snellen_fig}
\end{figure*}

In this tutorial, we have covered most of the techniques that are currently used to characterize the atmospheres of exoplanets at optical and infrared wavelengths.  Still, we expect that clever investigators will invent new techniques in the future.  Already, some are thinking of detecting exoplanetary signatures such as auroral emission from Earth-like planets \citep{luger17}, and the scope and depth of exoplanetary atmospheric characterization is likely to expand dramatically in the next decade.  

\acknowledgments
We thank the referee, Prof.~Kevin~Heng, for insightful comments that greatly improved this tutorial.  We also benefitted from conversations with Prof.~Heather~Knutson concerning the use of prioring orbital parameters when solving for secondary eclipse depths, and on other issues that allowed us to sharpen this tutorial.  Drs. Eric Agol, Mateo Brogi, Julien de Wit, Michael Line, Peter McCullough, Kevin Stevenson, and graduate students Yayaati Chachan, Brian Healy, and Victor Trees also provided valuable comments.  We thank the authors of the numerous figures that we have reproduced in this tutorial, and their publishers.  Figure~\ref{hat26} is reproduced by permission of the American Association for the Advancement of Scince (Science magazine).   Figure~\ref{brogi_fig} is reproduced by permission of Springer Publishing (Nature), and Figure~\ref{triaud_fig} is reproduced by permission of Oxford University Press (MNRAS).



\end{document}